\documentclass[preprintnumbers,nofootinbib,secnumarabic]{revtex4}

\usepackage{graphicx}
\usepackage{amsmath,amssymb,mathrsfs}
\usepackage{fancyhdr}
\usepackage{psfrag}
\usepackage{array,bbm}
\usepackage{url}
\usepackage{rotating}
\usepackage{multirow}
\usepackage{dsfont}
\usepackage{cancel}

\usepackage{epsfig}

\numberwithin{equation}{section}
\makeatletter       

\renewcommand{\p@subsection}{}

\makeatother

\newenvironment{Eqnarray}%
     {\arraycolsep 0.14em\begin{eqnarray}}{\end{eqnarray}}
\newcommand{\beqa}{\begin{Eqnarray}}
\newcommand{\eeqa}{\end{Eqnarray}}
\newcommand{\beq}{\begin{equation}}
\newcommand{\eeq}{\end{equation}}

\def\eq#1{eq.~(\ref{#1})}
\def\eqs#1#2{eqs.~(\ref{#1}) and (\ref{#2})}

\def\phm{\phantom{-}}
\def\half{\tfrac{1}{2}}
\def\centeron#1#2{{\setbox0=\hbox{#1}\setbox1=\hbox{#2}\ifdim
\wd1>\wd0\kern.5\wd1\kern-.5\wd0\fi
\copy0\kern-.5\wd0\kern-.5\wd1\copy1\ifdim\wd0>\wd1
\kern.5\wd0\kern-.5\wd1\fi}}
\def\ltap{\;\centeron{\raise.35ex\hbox{$<$}}{\lower.65ex\hbox{$\sim$}}\;}
\def\gtap{\;\centeron{\raise.35ex\hbox{$>$}}{\lower.65ex\hbox{$\sim$}}\;}

\def\lsim{\mathrel{\ltap}}
\def\wtil{\widetilde}
\def\anti{\overline}
\def\ur{U_R}
\def\dr{D_R}
\def\lt{\left}
\def\rt{\right}

\begin{document}
\preprint{SCIPP 12/14}
\title{Mass-degenerate Higgs bosons at 125~GeV in the
  Two-Higgs-Doublet Model}

\author{P.M.~Ferreira}
    \email[E-mail: ]{ferreira@cii.fc.ul.pt}
\affiliation{Instituto Superior de Engenharia de Lisboa,
	1959-007 Lisboa, Portugal}
\affiliation{Centro de F\'{\i}sica Te\'{o}rica e Computacional,
    Faculdade de Ci\^{e}ncias,
    Universidade de Lisboa,
    Av.\ Prof.\ Gama Pinto 2,
    1649-003 Lisboa, Portugal}
\author{Howard E.~Haber}
    \email[E-mail: ]{haber@scipp.ucsc.edu}
\affiliation{Santa Cruz Institute for Particle Physics,
    University of California,
    Santa Cruz, California 95064, USA}
\author{Rui Santos}
    \email[E-mail: ]{rsantos@cii.fc.ul.pt}
\affiliation{Instituto Superior de Engenharia de Lisboa,
	1959-007 Lisboa, Portugal}
\affiliation{Centro de F\'{\i}sica Te\'{o}rica e Computacional,
    Faculdade de Ci\^{e}ncias,
    Universidade de Lisboa,
    Av.\ Prof.\ Gama Pinto 2,
    1649-003 Lisboa, Portugal}
\author{Jo\~{a}o P.~Silva}
    \email[E-mail: ]{jpsilva@cftp.ist.utl.pt}
\affiliation{Instituto Superior de Engenharia de Lisboa,
	1959-007 Lisboa, Portugal}
\affiliation{Centro de F\'{\i}sica Te\'{o}rica de Part\'{\i}culas (CFTP),
    Instituto Superior T\'{e}cnico, Universidade T\'{e}cnica de Lisboa,
    1049-001 Lisboa, Portugal}

\date{\today}

\begin{abstract}
The analysis of the Higgs boson data by the ATLAS and CMS
Collaborations appears to exhibit an excess of $h\to\gamma\gamma$ events
above the Standard Model (SM) expectations; whereas no significant
excess is observed in $h\to ZZ^*\to\textrm{four lepton}$ events,
albeit with large statistical uncertainty due to the small data
sample.  These results (assuming they persist with further data) could
be explained by a pair of nearly mass-degenerate scalars, one
of which is a SM-like Higgs boson and the other is
a scalar with suppressed couplings to $W^+W^-$ and $ZZ$.  In the
two Higgs doublet model, the observed $\gamma\gamma$ and
$ZZ^*\to\textrm{four lepton}$ data can be reproduced
by an approximately degenerate CP-even ($h$) and CP-odd ($A$) Higgs boson
for values of $\sin(\beta-\alpha)$ near unity and
$0.7\lsim\tan\beta\lsim1$.  An enhanced $\gamma\gamma$ signal can also
arise in cases where $m_h \simeq m_H$, $m_H \simeq m_A$,
or $m_h \simeq m_H \simeq m_A$.
Since the $ZZ^*\to\textrm{four lepton}$ signal derives primarily from a
SM-like Higgs boson whereas the $\gamma\gamma$ signal receives contributions from
two (or more) nearly mass-degenerate states, one would expect
a slightly different invariant mass peak in the $ZZ^*\to\textrm{four lepton}$ and
$\gamma\gamma$ channels. The phenomenological consequences of such models can be
tested with additional Higgs data that will be collected at the LHC in the
near future.

\end{abstract}

\maketitle

%
\section{Introduction}
\label{sec-intro}

The ATLAS~\cite{ATLASHiggs} and CMS~\cite{CMSHiggs} collaborations
have recently announced the discovery of a new boson
with the properties approximating those
of the Standard Model Higgs boson ($h$)~\cite{hhg}.
The cleanest channels for the observation of the Higgs boson are via
the rare decays $h\to\gamma\gamma$ and $h\to ZZ^*\to\ell^+\ell^-
\ell^{\prime\,+}\ell^{\prime\,-}$ (where $\ell=e$ or $\mu$).
Evidence for Higgs boson production and decay is also seen in
$h\to WW^*\to \ell^+\nu\ell^-\overline{\nu}$.  Using this data,
both ATLAS and CMS have compared the observed values of $\sigma\times
{\rm BR}$ for the Higgs boson production and decay into the various
final states, relative to the corresponding Standard Model predictions.
Including both the 2011 and 2012 data sets, both collaboration observe
excesses of the $\gamma\gamma$ channels~\cite{ATLASgg,CMSgg},
whereas the $WW^*$ and $ZZ^*$
channels are consistent with Standard Model expectations.

Although the excess in the $\gamma\gamma$ channel is not yet
statistically significant,\footnote{The $\gamma\gamma$ excess observed by the ATLAS collaboration
corresponds to 2.4 standard deviations from the Standard Model Higgs boson signal
plus background hypothesis~\cite{ATLASgg}.} it is
instructive to consider mechanisms that could enhance the $\gamma\gamma$
channel while not affecting the Standard Model--like results for
$pp$ into vector boson pairs. Note that the Higgs boson couples
to $\gamma\gamma$ through a one-loop process, whereas
it
couples to vector bosons at tree level.   Thus, one could enhance the
$\gamma\gamma$ signal by proposing new charged particles beyond the
Standard Model that couple to the Higgs boson~\cite{manyrefs1}.  Such particles would
contribute in the loop amplitude for $h\to\gamma\gamma$ and could
potentially enhance the $h\to\gamma\gamma$ branching ratio.
However, in such a scenario one must check that the same particles
do not contribute significantly to the $gg\to h$ production
mechanism (which is also generated by a one-loop process).  Otherwise,
one might find that $\sigma\times {\rm BR}$ for Higgs production and
decay into $WW^*$ and $ZZ^*$ is enhanced or suppressed with respect to
the Standard Model, which is disfavored by the observed data.

Alternatively, one could enhance the $\gamma\gamma$ signal by increasing
the branching ratio for Higgs decay into $\gamma\gamma$
with respect to that of the Standard Model.  One way to
accomplish this is to reduce the corresponding Higgs branching ratios into
other decay channels.  The most effective mechanism of this type
is one where the dominant Higgs decay rate into $b\bar b$ pairs is
significantly reduced.  Specific examples can be found in Ref.~\cite{manyrefs2}.

In this paper, we investigate a third mechanism that can potentially
enhance the $\gamma\gamma$ signal while not affecting the $WW^*$ and
$ZZ^*$ signal.  For simplicity, we assume that the Higgs sector is
CP-conserving.  Suppose that there exist two nearly mass-degenerate
scalars, a CP-even state $h$ and a CP-odd state $A$.  If the
properties of $h$ are approximately given by those of the Standard
Model Higgs boson, then the observed $\sigma\times BR(h \to VV^*)$
[where $V=W$ or $Z$] should be consistent with Standard Model Higgs
production and decay, due to the fact that there are no tree-level
$A VV$ couplings.\footnote{An interesting caveat,
investigated in Ref.~\cite{logan}, is that there is
a very large enhancement of $A \to Z \gamma$,
feeding into the $A \to 4 \ell$ used experimentally
to detect  $Z Z^*$.
In that case,
only $A \to W W^*$ would be suppressed.}
On the other hand, the production cross section
for $gg\to A$ is typically larger than the
production cross section for $gg\to h$, and the
decay $A\to\gamma\gamma$ is mediated at one-loop by the top quark
loop.  Thus, if $m_{h}\simeq m_{A}$,
then both the $h$ and the $A$ will appear in the
$\gamma\gamma$ signal, thereby yielding an enhanced $\gamma\gamma$ signal.

The same mechanism of enhancing the $\gamma\gamma$ signal applies more generally to
any Higgs sector that only contains Higgs
doublets (and possibly Higgs singlets) with no higher Higgs
multiplets.  If there exists a CP-even state $h$ whose properties
are approximately given by those of the Standard
Model Higgs boson, then this one state will approximately saturate the sum rule for
$h VV$ couplings~\cite{sumrule}, in which case the couplings
of any other neutral Higgs boson (whether CP-even, CP-odd or an arbitrary mixture thereof)
to $VV$ will be highly suppressed.  In such a case, $h$ and a second scalar state
are approximately mass-degenerate, then both can contribute to
an enhanced $\gamma\gamma$ signal observed at the LHC, whereas only
$h$ production and decay yields a significant $ZZ^*\to\rm{four~leptons}$
signal.

In this paper, we consider the implications of the CP-conserving
two Higgs doublet extension of the Standard Model (2HDM).
There is already a substantial literature that analyzes the implications
of the present Higgs data in the framework of the 2HDM~\cite{sher,2HDMrefs}.
We shall address the question of whether the presence of nearly mass-degenerate
neutral scalars of the 2HDM can be responsible for an enhanced $\gamma\gamma$
signal.  Such a mechanism was first proposed in  Ref.~\cite{Gunion:2012gc} in
the context of the non-minimal supersymmetric extension of the
Standard Model (NMSSM), and was also recently applied in the context
of the 2HDM in an independent study~\cite{degenerate2}.

What is the origin of the approximate mass degeneracy of the two
near-degenerate Higgs
states?  In most cases, the near-degeneracy is accidental.  However,
one could imagine that a near-mass-degenerate CP-even/CP-odd scalar pair
might originate from a single complex scalar $\Phi^0$.
The presence of a global U(1) symmetry would then
yield identical masses for the real
and imaginary parts of $\Phi^0$.  If the explicit
breaking of the global U(1) symmetry is small (perhaps due to loop
corrections), the resulting CP-even and CP-odd scalars would end up as
approximate mass-degenerate states.  Whether the near-degeneracy is
accidental or the result of an approximate symmetry, it is important
to analyze the LHC Higgs data to determine if such a scenario
is realized in nature or can be excluded.
In Ref.~\cite{Gunion:2012he}, a set of diagnostic tools was developed
that can be used to determine whether mass-degenerate Higgs states are
present in the LHC Higgs data.

We begin our analysis by computing the $\sigma\times {\rm BR}$
for the production and decay of $h$ and $A$ separately, under the
assumption that $m_{h}\simeq m_{A}\simeq 125$~GeV and the couplings
of $h$ to $W^+W^-$ and $ZZ$ approximate the corresponding Standard Model
Higgs boson couplings.  We then determine the region of 2HDM parameter
space in which the sum of the $h$ and $A$ signals into
$\gamma\gamma$ match the enhanced rates suggested by the central
values measured by the ATLAS and CMS Collaborations.
We also check to see whether an excess in the $\gamma\gamma$
channel can be achieved with other mass degenerate scalars
($h$, $H$), ($H$, $A$) and ($h$, $H$, $A$).
Under the assumption that mass-degenerate
neutral Higgs states can be invoked to explain the excess in
the $\gamma\gamma$ channel, we predict which additional Higgs channels
must also exhibit deviations from their corresponding Standard Model
predictions.

\section{Setting up the 2HDM parameter scan}

The 2HDM consists of two hypercharge-one scalar doublet fields, denoted by
$\Phi_a$, where $a=1,2$~\cite{review}.
The Lagrangian of the most general 2HDM contains a scalar potential,
with two real and one complex scalar squared-mass term and four
real and three complex dimensionless self-couplings, and
the most general set of dimension-4
Higgs-fermion Yukawa interactions.  For example,
the most general Yukawa Lagrangian, in terms of the quark mass-eigenstate fields, is:
\beq \label{yuk}
-\mathscr{L}_{\rm Y}=\anti U_L \wtil\Phi_{a}^0{\eta^U_a} \ur +\overline
D_L K^\dagger\wtil\Phi_{a}^- {\eta^U_a}\ur
+\overline U_L K\Phi_a^+{\eta^{D\,\dagger}_{a}} \dr
+\overline D_L\Phi_a^0 {\eta^{D\,\dagger}_{a}}\dr+{\rm h.c.}\,,
\eeq
where $\wtil\Phi_{a}\equiv (\wtil\Phi^0\,,\,\wtil\Phi^-)
=i\sigma_2\Phi_{a}^*$ and $K$ is the CKM mixing matrix.
In \eq{yuk}, there is an implicit sum over the index $a=1,2$, and the $\eta^{U,D}$
are $3\times 3$ Yukawa coupling matrices.
However, such models generically possess
tree-level Higgs-mediated flavor-changing neutral currents (FCNCs), in conflict
with experimental data that requires FCNC interactions to be significantly
suppressed.

To avoid dangerous FCNC interactions in a natural way, we impose
a $\mathbb{Z}_2$ symmetry on the dimension-4 terms of the Higgs Lagrangian
in such a way that removes two of the four Yukawa coupling
matrices of \eq{yuk}.  The corresponding symmetry consists of choosing one of the
two Higgs fields to be odd under the $\mathbb{Z}_2$ symmetry.  The resulting
Higgs potential is given by:
\beqa
 V &=& m^2_{11}\Phi_1^\dagger\Phi_1+m^2_{22}\Phi_2^\dagger\Phi_2
   -\lt(m^2_{12}\Phi_1^\dagger\Phi_2+{\rm h.c.}\rt)
   +\half\lambda_1\lt(\Phi_1^\dagger\Phi_1\rt)^2
   +\half\lambda_2\lt(\Phi_2^\dagger\Phi_2\rt)^2 \nonumber \\
&&\qquad\qquad\qquad\qquad\qquad\qquad +\lambda_3\Phi_1^\dagger\Phi_1\Phi_2^\dagger\Phi_2
   +\lambda_4\Phi_1^\dagger\Phi_2\Phi_2^\dagger\Phi_1
   +\lt[\half\lambda_5\lt(\Phi_1^\dagger\Phi_2\rt)^2+{\rm h.c.}\rt]\,,\label{pot}
\eeqa
where we have allowed for a softly-broken $\mathbb{Z}_2$ symmetry due to
the presence of the
term proportional to $m_{12}^2$.  For simplicity, we shall assume that
the potentially complex
terms in \eq{pot}, $m_{12}^2$ and $\lambda_5$ are real, in which case the
Higgs potential
is CP-conserving.  In addition,
some of the fermion fields may also be odd under the $\mathbb{Z}_2$ symmetry,
depending
on which two of the four Yukawa coupling matrices are required to vanish.
Different choices
lead to different Yukawa interactions.  In this paper, we focus on two different
model choices,
known in the literature as Type-I~\cite{type1,hallwise} and
Type-II~\cite{type2,hallwise}.
In the Type-I 2HDM, $\eta_1^U=\eta_1^D=0$ in
\eq{yuk}, which means that all quarks and leptons couple exclusively to $\Phi_2$.
In the Type-II 2HDM, $\eta_1^U=\eta_2^D=0$ in \eq{yuk},
which means that the up-type quarks couple
exclusively to $\Phi_2$ and the down-type quarks and charged leptons
couple exclusively to
$\Phi_1$.

We assume that the parameters of the Higgs potential are chosen
such that the SU(2)$\times$U(1)
electroweak symmetry is broken to U(1) electromagnetism.
The neutral Higgs fields then
acquire vacuum expectation values, $\langle\Phi_a^0\rangle=v_a/\sqrt{2}$,
where $v_1^2+v_2^2=4m_W^2/g^2=(246~{\rm GeV})^2$
and $\tan\beta\equiv v_2/v_1$.
By convention, we choose $0\leq\beta\leq\half\pi$.
Expanding the Higgs potential about its minimum, we
then diagonalize the resulting scalar squared-mass matrices.
Three Goldstone boson states are eaten by the
$W^\pm$ and $Z$ gauge bosons, leaving five physical degrees of freedom:
a charged Higgs pair, $H^\pm$, two
CP-even neutral Higgs states, $h$ and $H$ (defined such that
$m_{h}\leq m_{H}$), and one CP-odd
neutral Higgs boson $A$.  In diagonalizing the CP-even neutral
Higgs squared-mass matrix, one also
obtains a CP-even Higgs mixing angle, $\alpha$.
By convention,
we take $|\alpha|\leq\pi/2$.
In the Type-I and Type-II 2HDM, the Higgs-fermion couplings are
flavor diagonal and depend on the
two angles $\alpha$ and $\beta$ as shown in
Table~\ref{tab:couplings}.
\begin{table}[h!]
\begin{center}
\begin{tabular}{|c|ccccccc|ccccccc|}
\hline
 & & \multicolumn{5}{c}{Type-I} &  \hspace{3ex}& &\multicolumn{5}{c}{Type-II} &\\
 & & $h$  & & $A$ & & $H$ & \hspace{3ex}& & $h$ & & $A$ & & $H$& \\
\hline
Up-type quarks  & &
$\cos{\alpha}/\sin{{\beta}}$   & &
$\phm\cot{\beta}$  & &
$\sin{\alpha}/\sin{{\beta}}$   &  \hspace{3ex} & &
$\phm\cos{\alpha}/\sin{{\beta}}$   & &
$\cot{\beta}$ & &
$\sin{\alpha}/\sin{{\beta}}$ &\\
Down-type quarks and charged leptons & &
$\cos{\alpha}/\sin{{\beta}}$   & &
$- \cot{\beta}$  & &
$\sin{\alpha}/\sin{{\beta}}$  &  \hspace{3ex} & &
$- \sin{\alpha}/\cos{{\beta}}$   & &
$\tan{\beta}$ & &
$\cos{\alpha}/\cos{{\beta}}$ & \\
\hline
\end{tabular}
\end{center}
\caption{Couplings of the fermions to the lighter and heavier
CP-even scalars ($h$ and $H$),
and the CP-odd scalar ($A$).
\label{tab:couplings}}
\end{table}

The aim of this paper is to scan the Type-I and Type-II 2HDM
parameter spaces allowing the $\gamma\gamma$ signal resulting from the production of
two near-mass-degenerate Higgs states
to be enhanced above the Standard Model (SM) rate, while assuming that the
corresponding $ZZ^*\to 4$~leptons (and $WW^*$) signal
is approximately given by the corresponding SM rate.
In general,  given a final state $f$,
we define $R_f$ as the ratio of the number of events
predicted in the 2HDM with near-mass-degenerate Higgs
states to that obtained in the SM:
\begin{equation}
R_f = \frac{N(pp \to f)_{\rm 2HDM}}{N(pp \to f)_{\rm SM}}.
\end{equation}
Denoting the two near-mass-degenerate Higgs states by $h_1$ and $h_2$,
we obtain
\begin{equation}
R_f = R^{h_1}_f + R^{h_2}_f,
\end{equation}
where
\begin{equation} \label{Sratio}
R^S_f =
\frac{\sigma(pp \to S)_{\textrm{2HDM}}\
\textrm{BR}(S \to f)_{\textrm{2HDM}}}{
\sigma(pp \to h_{\textrm{SM}})\ \textrm{BR}(h_{\textrm{SM}} \to f)},
\end{equation}
for $S=h_1$, $h_2$ and $\sigma$ is the Higgs production
cross section, BR the branching ratio,
and $h_{\textrm{SM}}$ is the SM Higgs boson.
In our analysis,
we include all Higgs production mechanisms,
namely,
gluon-gluon fusion using HIGLU at NLO~\cite{Spira:1995mt},
vector boson fusion (VBF)~\cite{LHCHiggs}, Higgs production in
association with either $W$, $Z$  or
$t\bar{t}$~\cite{LHCHiggs},
and  $b \bar{b}$ fusion~\cite{Harlander:2003ai}.

Since we do not expect Higgs coupling measurements at the LHC
to be better than about $20\%$
after the full 2012 data set is analyzed, we have performed our
2HDM scans under the
assumption that
\beq \label{rzz}
0.8 < R_{ZZ} < 1.2\,.
\eeq
In the case where $h_1=h$ and $h_2=A$, the
constraint imposed by our $R_{ZZ}$ assumption implies that
the $ZZ^*\to 4$~leptons signal
is due almost entirely to the production and decay of $h$,
due to the absence of a
tree-level $A ZZ$ coupling.
That is, the properties of the $h$ are SM-like.
Remarkably, the same conclusions emerge
in the case where both $h_1$ and $h_2$ are CP-even states.
In this case, we find that one of the
two CP-even states is SM-like, where the other has suppressed
tree-level couplings to $ZZ$.

We perform our 2HDM scans separately for the Type-I and
Type-II Higgs-fermion couplings, by scanning over
the neutral and charged Higgs masses, the angles $\alpha$ and $\beta$ and
the soft-breaking parameter $m_{12}^2$,
subject to the constraint of \eq{rzz} and the near mass-degeneracy of two
(or three) neutral Higgs bosons.   If the two nearly mass-degenerate Higgs
states are CP-even, then we shall assume that the mass difference of
these states is large enough (compared to the corresponding Higgs boson widths)
so that we can neglect possible
interference effects in the production and decay process.\footnote{In the case of a mass-degenerate CP-even/CP-odd Higgs pair, there
is no interference due to CP invariance.}  This is not a significant
constraint on our analysis, as the Higgs widths are significantly smaller than the
experimental mass resolution of the ATLAS and CMS experiments.

In addition,
the allowed points of our scans must also satisfy the known
indirect experimental constraints on the Type-I and Type-II 2HDM.
These constraints are described in Section~\ref{sec:constraints}.
After imposing these requirements,
we identify the surviving parameter regimes where the $\gamma\gamma$
signal is enhanced, and discuss
additional Higgs phenomena that must be observed in order to confirm
the scenarios advocated in this paper.

\section{Constraints on the 2HDM parameters}
\label{sec:constraints}

In this section, we summarize the indirect experimental constraints on the CP-conserving
2HDM with Type-I and Type-II Higgs-fermion
Yukawa couplings, respectively.
First, we scan over the parameter
space subject
to the following three constraints: the Higgs scalar potential is
(i)~bounded from below~\cite{vac1};
(ii)~satisfies tree-level unitarity~\cite{unitarity}; and
(iii)~is consistent with constraints from the Peskin-Takeuchi $S$
and $T$ parameters~\cite{Peskin:1991sw,STHiggs}
as derived from electroweak precision
observables~\cite{lepewwg,gfitter1,gfitter2}.
Additional indirect 2HDM constraints arise from charged Higgs
exchange contributions
to processes involving the $b$ quark and $\tau$-lepton.
Such processes yield constraints in
the $m_{H^\pm}$--$\tan\beta$ plane.  Consequently, if the rates
for these processes are consistent
with SM expectations, then the corresponding 2HDM constraints are
(typically) relaxed in the limit
of large $m_{H^\pm}$ due to the decoupling properties of the
2HDM~\cite{decoupling}.
At present, there is an anomaly observed by the BaBar collaboration
in the rate for $\overline{B}\to D^{(*)}\tau^-\overline{\nu}_\tau$
which deviates by 3.4~$\sigma$ (when $D$ and $D^*$ final states are
combined) from the SM prediction~\cite{Lees:2012xj}.
The observed deviation cannot be explained by a charged Higgs boson
in the Type-II 2HDM.
Indeed the analysis of Ref.~\cite{Lees:2012xj} excludes the Type-II
2HDM for any value of $\tan\beta/m_{H^\pm}$
at the $99.8\%$ CL.  Until this observation is independently confirmed
by the BELLE collaboration, we will
not include this result among our 2HDM constraints.

A number of other observables in $B$ physics provide constraints
in the $m_{H^\pm}$--$\tan\beta$ parameter plane.
The most precise SM prediction for
$R_b\equiv\Gamma(Z\to b\bar{b})/\Gamma(Z\to{\rm hadrons})$,
including electroweak two-loop
and QCD three-loop corrections, deviates by two standard deviations
from the experimentally measured value~\cite{Freitas:2012sy}.
The inclusion
of 2HDM contributions could alleviate this discrepancy in particular
regions of the parameter
space~\cite{Denner:1991ie,Boulware:1991vp,Grant:1994ak,Haber:1999zh}.
Another constraining observable is the decay rate for $b\to s\gamma$.
The theoretical prediction for $b \to s \gamma$ in the SM
has been performed to order
$\mathcal{O}(\alpha_s ^2)$~\cite{Misiak:2006zs}
and is in good agreement
with the experimental result~\cite{Asner:2010qj}.
However, due to theoretical uncertainties coming mainly
from higher order QCD corrections and the uncertainties in the CKM matrix element,
there is still some room for 2HDM contributions.
Other $B$ observables such as $B$--$\bar{B}$ mixing data and the rate for
$B^+ \to \tau^+ \nu_\tau$ also provide constraints on the 2HDM parameter space.

In the Type-II 2HDM, $R_b$ provides the most restrictive bound
in the $m_{H^\pm}$--$\tan \beta$ plane in the small $\tan \beta$
region, as shown in Refs.~\cite{Haber:1999zh,gfitter1}.
However, the strongest bound on the charged Higgs
mass derives from the $b \to s \gamma$ measurement, which
yields $m_{H^\pm} \gtrsim 360$ GeV at the $95\%$ CL almost
independently of the value of $\tan \beta$~\cite{Mahm}.
In contrast,
$B$--$\bar{B}$ mixing yields less constraining bounds in
the $m_{H^\pm}$--$\tan \beta$ plane,
whereas $B^+ \to \tau^+ \nu_\tau$ constrains mainly
large values of $\tan \beta$ for small charged Higgs masses~\cite{BB}.
In the Type-I 2HDM, the most restrictive bound in the
$m_{H^\pm}$--$\tan \beta$ plane is due to
$b \to s \gamma$~\cite{BB}.  In contrast to the Type-II model,
there is no strong bound (independent of the value of $\tan \beta$)
on the charged Higgs mass.  Finally we note that
all other experimental bounds on the Type-I 2HDM from
$B$-physics observables were shown to be less
restrictive~\cite{BB,gfitter1}.

With the exception of $B\to \tau^+\nu_\tau$ and
$\overline{B}\to D^{(*)}\tau^-\overline{\nu}_\tau$
(where charged Higgs exchange contributes at tree-level),
the 2HDM contributions
to $B$ physics observables arise via one-loop radiative corrections.
For such observables,
it is always possible that constraints on the 2HDM parameter space
could be relaxed due to
cancellations in the loop from other sources of new physics.
One of the well-known examples of
this phenomenon is the partial cancellation of the charged Higgs
loop and the chargino loop
contributions to  $b \to s \gamma$ in the MSSM~\cite{bsgmssm}.
Thus, to be flexible in our presentation, we shall present results
of our 2HDM scans
with and without the bounds from $B$-physics
and $R_b$.  In the former case, the $95\%$ CL bounds from $B$-physics
and $R_b$ are employed.
We will denote our parameter scan as
``constrained,'' when \textit{all} experimental and theoretical bounds
are considered, and ``unconstrained'' when
all except the $R_b$ and $B$-physics bounds are taken into account.

\section{Degenerate $h$ and $A$}

We begin by fixing $m_A = m_h = 125\, \textrm{GeV}$.
We scan over 2HDM parameters subject to \eq{rzz}. As a consequence,
the $h$ couplings to $W^+W^-$, $ZZ$ and $t\bar{t}$ cannot
differ much from their SM values.
The $\gamma \gamma$ signal is enhanced if the contribution
from the $A$ is large;
this can be achieved for values of $\tan{\beta}\lsim 1$.
Thus, we focus our analysis on the region
$0.5 < \tan\beta < 2$.
Since $A$ does not couple to $Z Z$,
the square of the $h ZZ$ coupling, {\em i.e.} $\sin^2{(\beta - \alpha)}$,
is constrained to be near its SM value.
In the $\tan\beta$ parameter regime of interest,
we find that $0.7 \lsim \sin^2{(\beta - \alpha)} < 1$
for both Type-I and Type-II Higgs-fermion Yukawa
couplings.\footnote{This result
is not surprising. For values of $\tan\beta\sim 1$,
the production cross-section for $h$ production is SM-like.
In light of the constraint of \eq{rzz},
it follows that the $hZZ$ coupling should be approximately SM-like,
which implies a
value of $\sin(\beta-\alpha)$ close to 1~\cite{hhg,decoupling}.}

By convention (and without loss of generality), we define $\alpha$
and $\beta$ such that $-\half\pi\leq\alpha\leq\half\pi$ and
$0\leq\beta\leq\half\pi$.
Thus, for $0.5<\tan\beta<2$ it follows
that $0.84\lsim\sin(\beta-\alpha)<1$.
In particular,
negative values of $\sin(\beta-\alpha)$ near
$-1$ are not permitted.
In this parameter regime,
$|\cos(\beta-\alpha)|\lsim 0.55$ with both signs allowed.
Given these constraints,
the range of possible values of $\alpha$ is also constrained.
Using the trigonometric identities,
\beqa
\sin\alpha&=&\sin\beta\cos(\beta-\alpha)-\cos\beta\sin(\beta-\alpha)\,,\label{sa}\\
\cos\alpha&=&\cos\beta\cos(\beta-\alpha)+\sin\beta\sin(\beta-\alpha)\,,\label{ca}
\eeqa
one can easily determine the allowed range of possible $\alpha$ values in our scans.

In our 2HDM parameter scan, the mass of the heavier CP-even scalar ($m_H$)
is kept between 200 and 1000 GeV,
and $\alpha$ is allowed to vary subject to the constraints implicit in
\eqs{sa}{ca}.  As discussed in Section~\ref{sec:constraints},
we impose the requirements of a scalar potential that is
bounded from below and satisfies
unitarity, and apply the constraints from precision
electroweak observables.  We also vary the charged Higgs mass between
500 and 1000 GeV subject to all the constraints
discussed in Section~\ref{sec:constraints}.
After imposing all the relevant constraints, we note
that the one-loop diagrams mediated by the
charged Higgs boson contribute very little to
$pp \to (h, A) \to \gamma \gamma$.

\subsection{Type-I 2HDM}

In the simulation of a mass-degenerate $h$, $A$
pair in the Type-I 2HDM,
we find $\sin{(\beta - \alpha)} > 0.88$
for $\tan{\beta} \sim 0.8$
and $\sin{(\beta - \alpha)} > 0.93$
for $\tan{\beta} \sim 1.6$.
These results are easy to understand in the approximation where
Higgs production is due exclusively to gluon-gluon fusion
via the top quark loop,
and the total Higgs width is well approximated by $\Gamma(h \to b \bar{b})$.
In this simplified scenario,
\begin{equation}
R_{ZZ} =
\frac{\cos^2{\alpha}}{\sin^2{\beta}}
\sin^2{(\alpha - \beta)} \frac{\sin^2{\beta}}{\cos^2{\alpha}}
=
\sin^2{(\alpha - \beta)}.
\label{RZZ_I_simplified}
\end{equation}
Using the Type-I couplings of Table~\ref{tab:couplings}, the first factor derives from the square of the $t\bar{t}h$ coupling
that governs the gluon-gluon fusion cross section,
the second from $\Gamma(h \to Z Z)$
and the third from
$\Gamma_{\textrm{total}} \sim \Gamma(h \to b \bar{b})$
which appears in the denominator.
Thus $\sin^2{(\alpha - \beta)} = R_{ZZ} > 0.8$
as a result of the constraint imposed by \eq{rzz}.
This precludes small values for $\sin^2{(\alpha - \beta)}$,
and we expect
SM-like couplings of $h$ to gauge bosons.

\begin{figure}[h!]
\centering
\includegraphics[width=3.5in,angle=0]{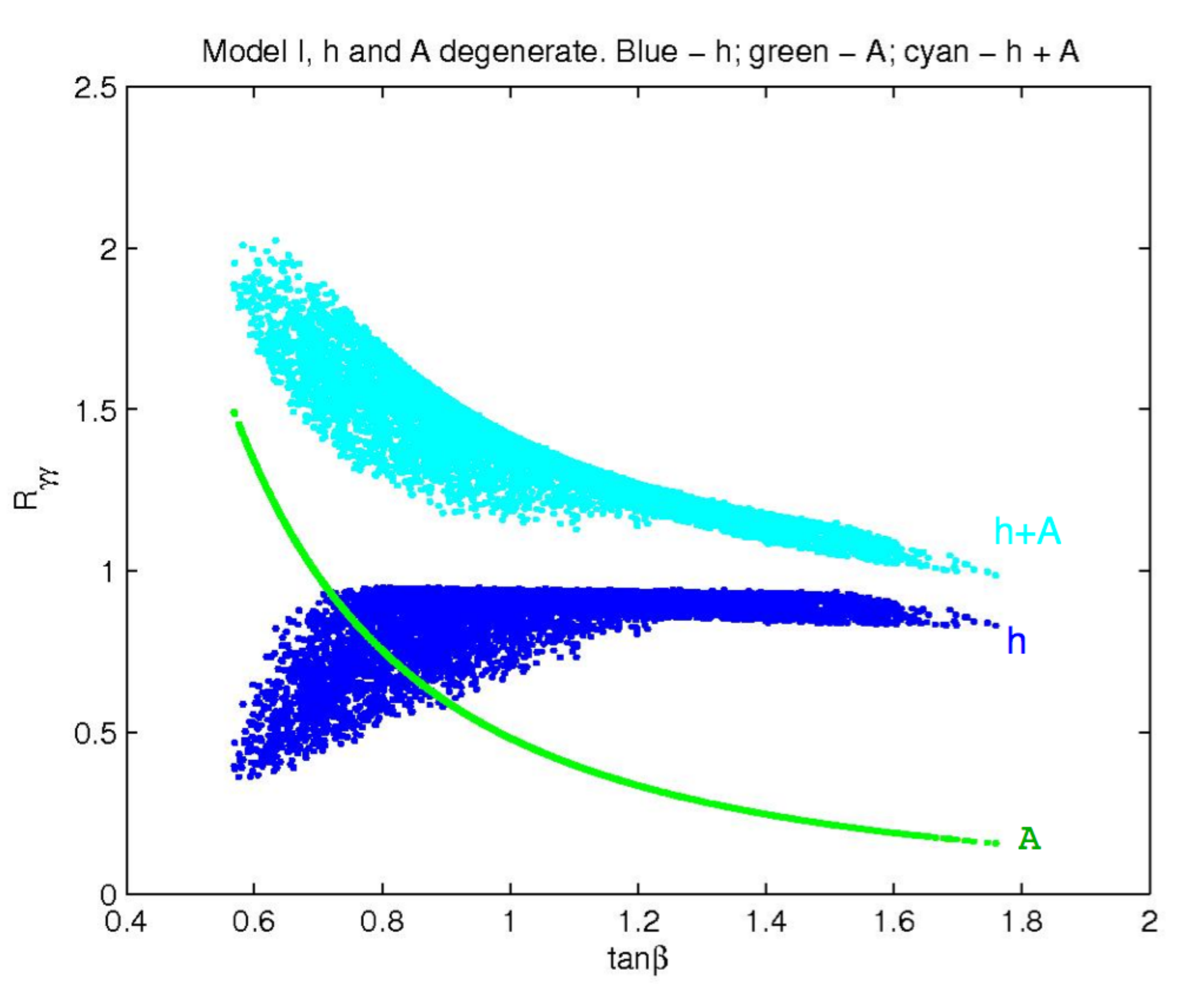}
\hspace{-.3cm}
\includegraphics[width=3.5in,angle=0]{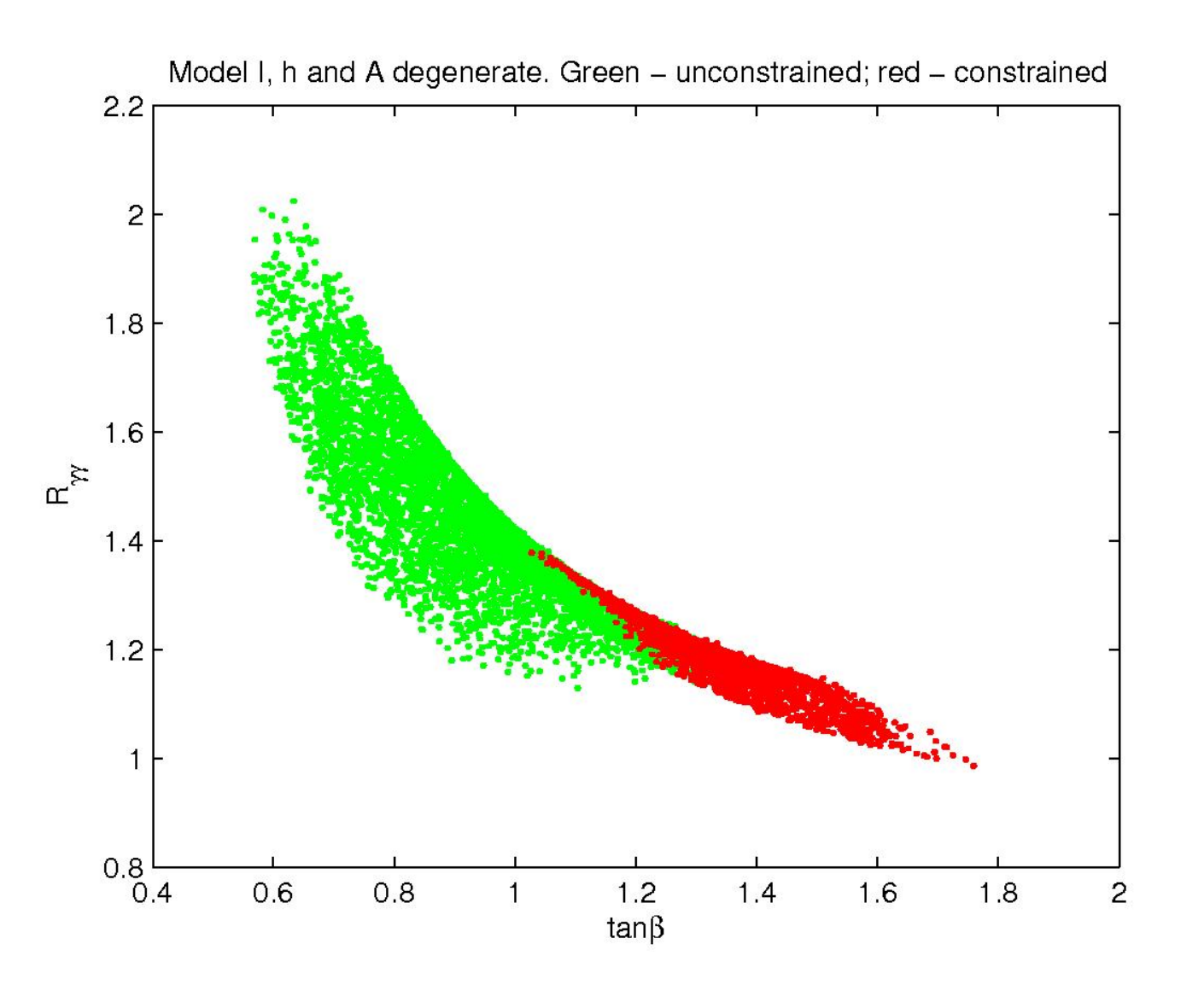}
\caption{Left panel: $R_{\gamma \gamma}$ as a function of $\tan{\beta}$ for
$h$ (blue/black),
$A$ (green/gray),
and the total observable rate (cyan/light-gray),
obtained by summing the rates with intermediate $h$ and $A$,
for the unconstrained scenario.
Right panel:  Total rate for $R_{\gamma \gamma}$ as a function of $\tan{\beta}$
for the constrained (red/black) and unconstrained (green/gray) scenarios.}
\label{fig:hA_modI_phph}
\end{figure}

We have generated a large sample of points in parameter space
satisfying all constraints.
The results for $R_{\gamma \gamma}$ as a function of $\tan{\beta}$
are presented in Fig.~\ref{fig:hA_modI_phph} for the
unconstrained scenario (left panel) and the
constrained scenario (right panel).
We note that $R^h_{\gamma \gamma}$ lies below
unity regardless of $\tan{\beta}$; the wide region of the distributed
points reflects the variation as the CP-even mixing angle $\alpha$ is scanned.
This result was
previously obtained in Ref.~\cite{sher} in the limit of
$m_A \gg m_h$,
and in Ref.~\cite{Barroso:2012wz},
allowing for the mixing with the heavier CP-even and CP-odd scalars.
In contrast,
$R^A_{\gamma \gamma}$ increases as $\tan{\beta}$ decreases,
as expected,
due to the increase in the $At\bar{t}$ coupling
[cf.~Table~\ref{tab:couplings}].  Its value is uniquely determined by
$\tan\beta$, as there is no $\alpha$ dependence in the coupling of $A$
to fermions.  As a result, the observed
rate for $pp \to (h, A) \to \gamma \gamma$ normalized to the corresponding SM rate can take values
as large as $2$,
for $\tan{\beta} \sim 0.6$. However, constraints from $B$-physics
restrict the normalized rate for $pp \to (h, A) \to \gamma \gamma$ to
a maximal value of approximately $1.4$,
for $\tan{\beta} \sim 1$.

Since $A$ does not couple to $VV$,
the $\gamma \gamma$ decays detected in VBF production can only
be due to the $h$ intermediate state.
The ATLAS and CMS experiments can
isolate Higgs signal events with a set of additional criteria (e.g.~events with two
forward jets with certain transverse momentum cuts and a central jet veto)
which they designate as VBF Higgs events.
In practice the experimental VBF
Higgs events have significant
contamination\footnote{The ATLAS and CMS collaborations
quote contamination rates for the gluon--gluon fusion
Higgs events of roughly $30\%$~\cite{ATLASHiggs,CMSHiggs}, although
in practice this number has a rather large error bar.}
of Higgs events produced by gluon--gluon fusion
with the subsequent radiation of two additional jets.
Nevertheless, in this
work we find it convenient to define a theoretical quantity,
\beq
R_{\gamma\gamma}^{\rm VBF}=
\frac{\sigma(pp\to VV \to h)_{\textrm{2HDM}}
\ \textrm{BR}(h \to \gamma\gamma)_{\textrm{2HDM}}}{
\sigma(pp\to VV\to h_{\textrm{SM}})
\ \textrm{BR}(h_{\textrm{SM}} \to \gamma\gamma)}\,,
\eeq
which would be appropriate if VBF Higgs events could be identified
with no contamination.
This will prove sufficient for our purposes in this initial study.
Likewise, we shall denote
$R_{\gamma\gamma}\equiv R^h_{\gamma\gamma}+R^A_{\gamma\gamma}$
following the definitions
given in \eq{Sratio}.

\begin{figure}[t!]
\epsfysize=7cm
\centerline{\epsfbox{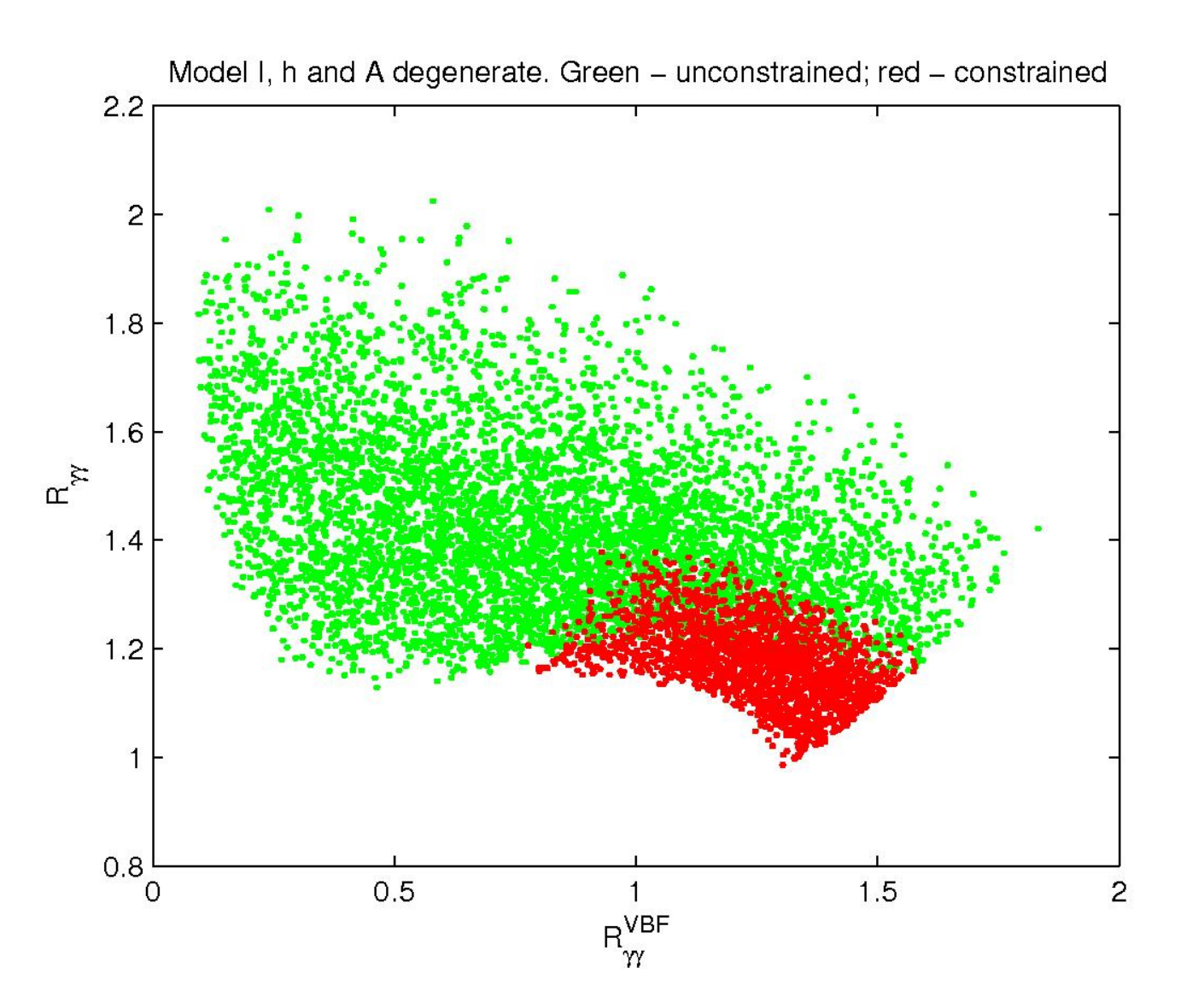}}
\caption{Allowed region in the
$R^{\textrm{VBF}}_{\gamma \gamma} - R_{\gamma \gamma}$
plane with (red/black) and without (green/gray) the $B$-physics constraints.}
\label{fig:vbf_vs_totI}
\end{figure}

Fig.~\ref{fig:vbf_vs_totI} shows the allowed region in the
$R^{\textrm{VBF}}_{\gamma \gamma}$--$R_{\gamma \gamma}$
plane, for our set of points, for the constrained (red/black)
and unconstrained (green/gray) scenarios.
In the unconstrained scenario,
contrary to what happens for the total $R^h_{\gamma \gamma}$,
the $h$ intermediate state can induce values of
$R^{\textrm{VBF}}_{\gamma \gamma}$ larger than unity.
The dispersion of points shows that,
in the region of parameter space that we have studied,
a comparison between $R^{\textrm{VBF}}_{\gamma \gamma}$ and
$R_{\gamma \gamma}$ is unlikely to  allow
an exclusion of the Type-I 2HDM.
Indeed,
for $R_{\gamma \gamma} \sim 1.5$,
this model allows for any value of
 $R^{\textrm{VBF}}_{\gamma \gamma}$ between $0.2$ and $1.7$.
Only if $R_{\gamma \gamma} \sim 2.0$
can we exclude values of $R^{\textrm{VBF}}_{\gamma \gamma}$
above unity.

In the constrained scenario,
$R^{\textrm{VBF}}_{\gamma \gamma}$
is now between $0.7$ and $1.6$,
while $R_{\gamma \gamma}$ lies
in the range from 1 to 1.4. The allowed values (red/black) clearly
show that an enhancement in $R_{\gamma \gamma}$
can only be achieved for $R^{\textrm{VBF}}_{\gamma \gamma}$
close to 1. Conversely, large values of $R^{\textrm{VBF}}_{\gamma \gamma}$
are attained only for a SM-like $R_{\gamma \gamma}$.
Naively, it seems puzzling that values of
$R_{\gamma\gamma}^{\rm VBF}$ above 1 are possible.
After all, this quantity is only sensitive to $h$
production and decay, and we are
assuming that the $hVV$ coupling is close to its SM value.
A closer examination of the parameter scan reveals that the
range of allowed $\alpha$ is rather limited,
$-0.9\lsim\sin\alpha\lsim -0.4$.
For the values of $\sin\alpha$ close to $-1$,
the Type-I couplings of $h$ to fermion pairs
are suppressed relative to the
corresponding SM couplings.
Consequently, the partial width of $h\to b\bar{b}$ is
reduced and so the corresponding branching ratio for $h\to\gamma\gamma$ is enhanced
relative to its SM value.\footnote{There is a small additional enhancement to the
$h\gamma\gamma$ partial width since the contribution of the top quark loop (which
negatively interferes with the dominant $W^\pm$ loop) is also reduced.}

\begin{figure}[h!]
\centering
\includegraphics[width=3.5in,angle=0]{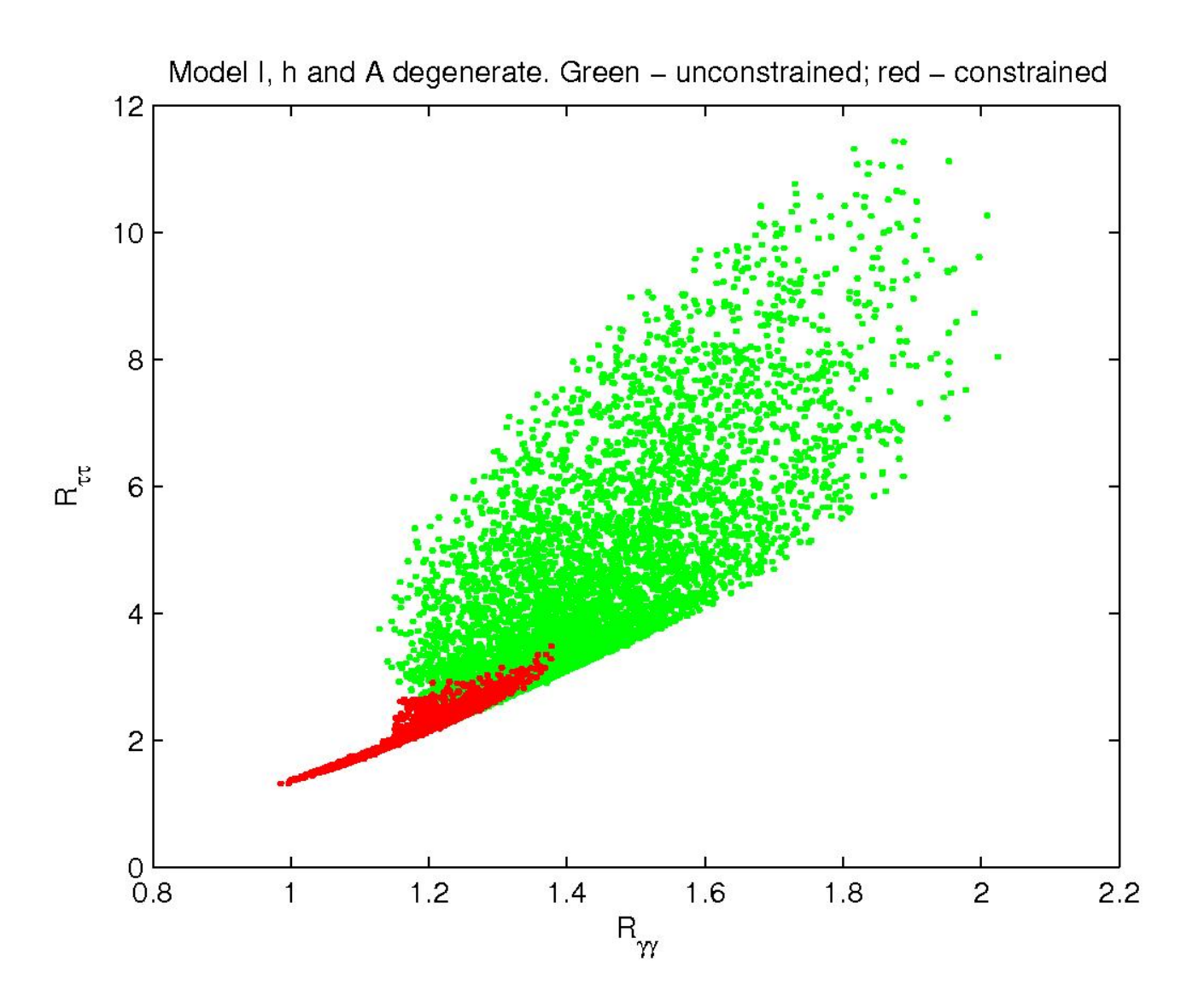}
\hspace{-.3cm}
\includegraphics[width=3.5in,angle=0]{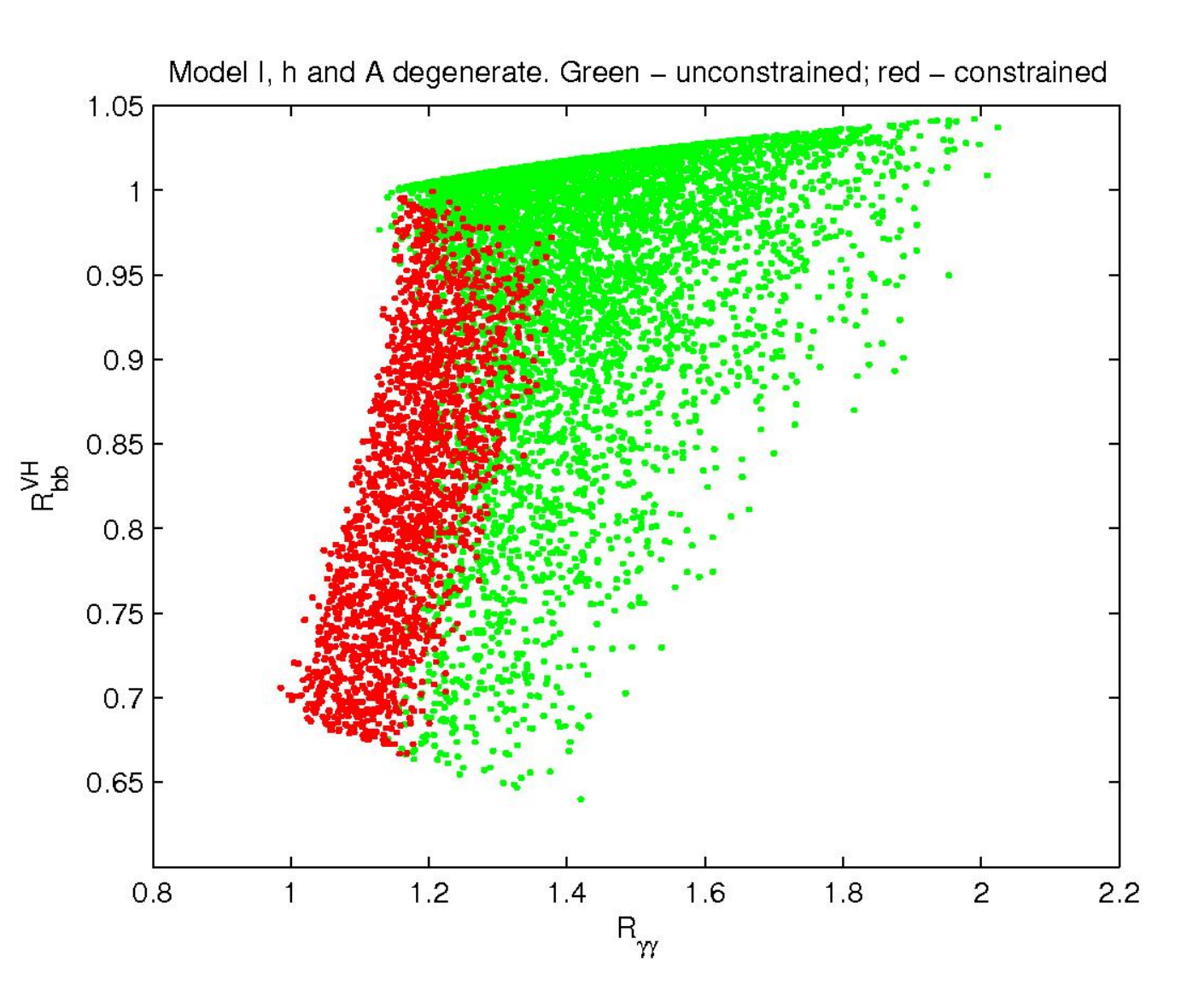}
\caption{Left panel: Total $R_{\tau\tau}$ ($h$ and $A$ summed)
as a function of $R_{\gamma \gamma}$
for the constrained (red/black) and unconstrained (green/gray) scenarios.
QCD corrections aside,
this figure also holds approximately for $R_{bb}$.
Right panel: $R^\textrm{VH}_{bb}$ ($h$ and $A$ summed)
as a function of $R_{\gamma \gamma}$
for the constrained (red/black) and unconstrained (green/gray) scenarios.}
\label{fig:phph_vs_bbI}
\end{figure}
The left panel of Fig.~\ref{fig:phph_vs_bbI} exhibits our results
for the inclusive $\tau\tau$ final state,
summing over all production mechanisms,
for the constrained (red/black) and unconstrained (green/gray) scenarios.
Note that both $R_{\tau\tau}$ and $R_{\gamma\gamma}$ include contributions
from $h$ \textit{and} $A$.
For the unconstrained scenario, we see that if
$R_{\gamma \gamma} \sim 1.5$ then
the total $R_{\tau\tau}$ contribution cannot be smaller than about 3.5 and can be as large as 8.
When the $B$-physics constraints are included the main
difference is again that $R_{\gamma \gamma} \lesssim 1.4$, whereas
$R_{\tau\tau}$ lies in a very narrow band heavily dependent
on the particular value of $R_{\gamma \gamma}$ but always
below about 3.  The correlation between the enhanced $\gamma\gamma$ and $\tau\tau$ signals is
a noteworthy prediction for this scenario.

There are only very slight differences between $R_{\tau \tau}$
and $R_{bb}$, due to QCD corrections that distinguish the two processes,
so the left panel of Fig.~\ref{fig:phph_vs_bbI} would apply as well to a hypothetical measurement
of the inclusive $b\bar{b}$ final state.  Unfortunately,
the detection of $h \to b \bar{b}$ via gluon-gluon production at the LHC
is swamped by background and is not possible in the inclusive mode.
However, both the ATLAS and CMS collaborations expect some sensitivity
to the $b \bar{b}$ final state in the production
of the Higgs boson in association with a $W^\pm$ or $Z$ (where leptonic decays of
the vector bosons can be used to tag the event).
The right panel of Fig.~\ref{fig:phph_vs_bbI} exhibits our results
for the $b \bar{b}$ final state obtained through $VH$ production,
$R^{\textrm{VH}}_{bb}$, with (red/black) and without (green/gray) the
$B$-physics constraints.
Since $A$ is not produced by this mechanism,
we only get a contribution from the SM-like $h$
and the enhancement in the $b\bar{b}$ channel disappears.
Furthermore, in the constrained scenario,
even an enhancement in
$R_{\gamma \gamma}$ corresponds to
$R^{\textrm{VH}}_{bb}$ below 1.


\subsection{Type-II 2HDM}

In the analysis of a mass-degenerate $h$, $A$ pair in the Type-II 2HDM,
we have generated a large sample of points satisfying all constraints.
We again anticipate that
$\sin(\beta-\alpha)$ should be near 1.
Applying the same simplified
scenario that yielded \eq{RZZ_I_simplified} for the
Type-I 2HDM, we now obtain
\begin{equation}
R_{ZZ} =
\frac{\cos^2{\alpha}}{\sin^2{\beta}}
\sin^2{(\alpha - \beta)} \frac{\cos^2{\beta}}{\sin^2{\alpha}}\,,
\label{RZZ_II_simplified}
\end{equation}
where we have employed the Type-II couplings of Table~\ref{tab:couplings}.
Assuming that $\cos(\beta-\alpha)$ is small, we can expand in this small
quantity by making use of
\beq
\frac{\cos^2{\alpha}}{\sin^2{\beta}}\cdot
\frac{\cos^2{\beta}}{\sin^2{\alpha}}=
\left[\frac{\sin(\beta-\alpha)+\cot\beta\cos(\beta-\alpha)}{
\sin(\beta-\alpha)-\tan\beta\cos(\beta-\alpha)}\right]^2
=1+\frac{4\cos(\beta-\alpha)}{\sin 2\beta}+\mathcal{O}\bigl(\cos^2(\beta-\alpha)\bigr)\,.
\eeq
Inserting this result into \eq{RZZ_II_simplified} yields
\beq \label{cbma}
\cos(\beta-\alpha)\simeq \tfrac{1}{4}\sin 2\beta(R_{ZZ}-1)\,.
\eeq
In light of \eq{rzz}, the assumption that $\cos(\beta-\alpha)$ is small is justified.
In our simulation we find that the constraint imposed by \eq{rzz}
leads to  $\sin{(\beta - \alpha)} > 0.996$, which is an even tighter restriction than
suggested by \eq{cbma}.

\begin{figure}[b!]
\epsfysize=7cm
\centerline{\epsfbox{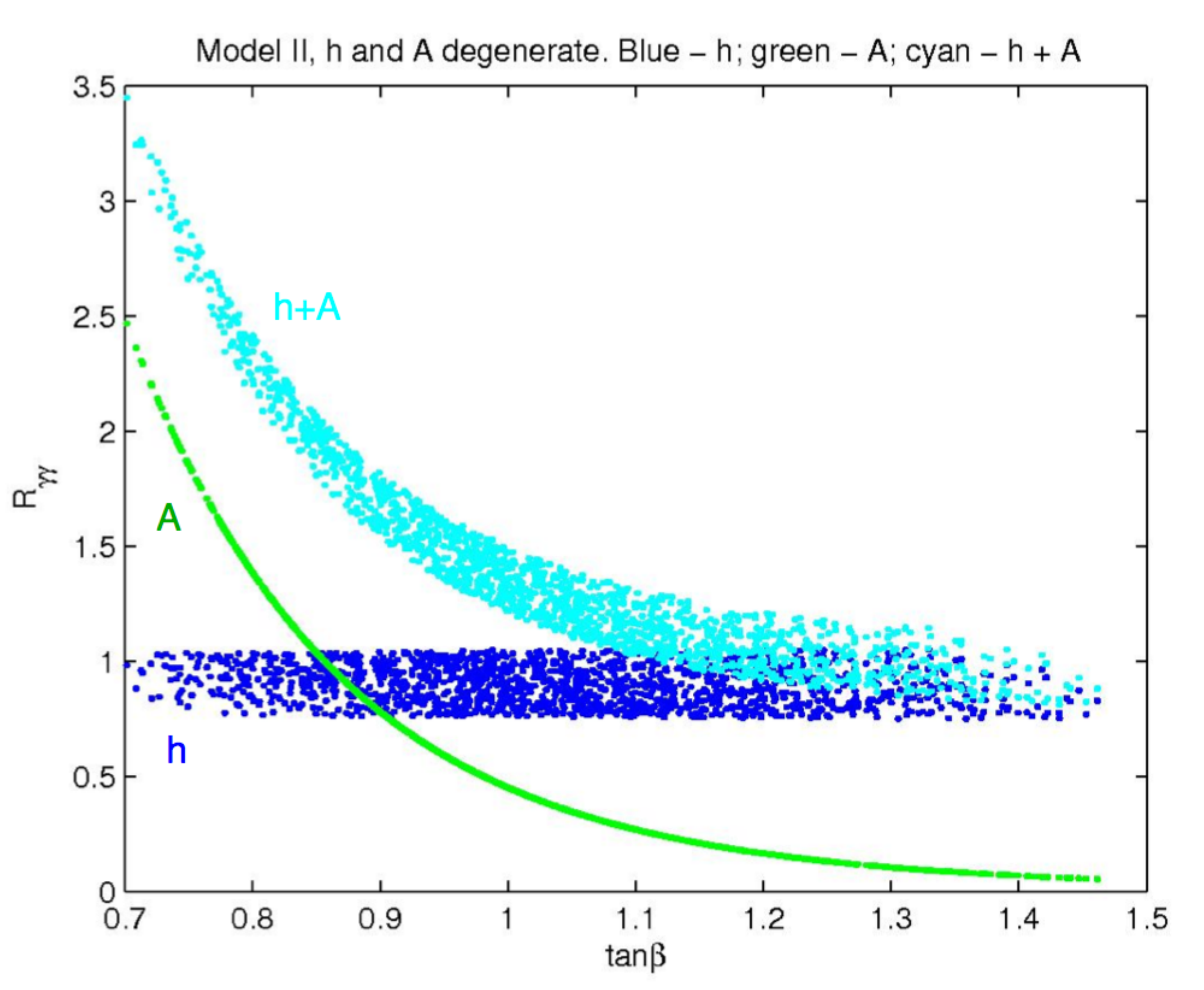}}
\caption{$R_{\gamma \gamma}$ as a function of $\tan{\beta}$ for
$h$ (blue/black),
$A$ (green/gray),
and the total observable rate,
obtained by summing the rates with intermediate $h$ and $A$ (cyan/light-gray).}
\label{fig:hA_modII_phph}
\end{figure}
Fig.~\ref{fig:hA_modII_phph}
shows $R_{\gamma \gamma}$ as a function of $\tan{\beta}$
for the Type-II 2HDM.
Here the $h$ contribution by itself can only reach unity and, as in the Type-I 2HDM,
the contribution from $A$ becomes dominant for low
$\tan{\beta}$ values.
The total $R_{\gamma \gamma}$ can be as large as $3.5$ for
$\tan{\beta} \sim 0.7$.
Moreover, the requirement that $0.8 < R_{ZZ} < 1.2$ leads to the
exclusion of any point with $\tan{\beta} < 0.7$ in our simulation.
Because the most important constraint for the Type-II 2HDM is the one from $R_b$
and since the charged Higgs mass is varied from 500 to 1000 GeV, $\tan{\beta}$
will not be affected by any further constraints from $B$ physics. This means that all results presented
for this Type-II scenario already incorporate all available experimental
and theoretical constraints.

\begin{figure}[h!]
\centering
\includegraphics[width=3.5in,angle=0]{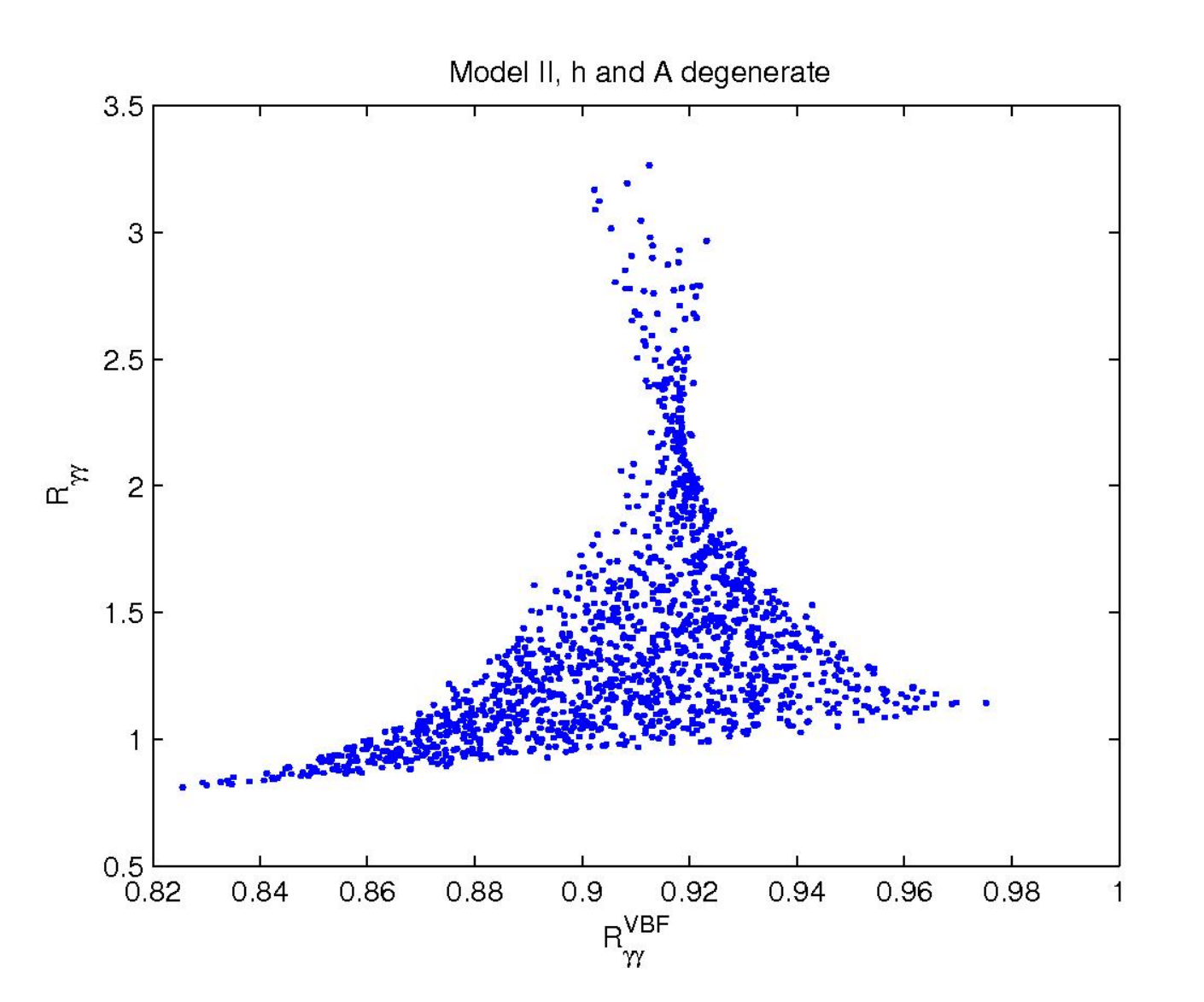}
\hspace{-.3cm}
\includegraphics[width=3.5in,angle=0]{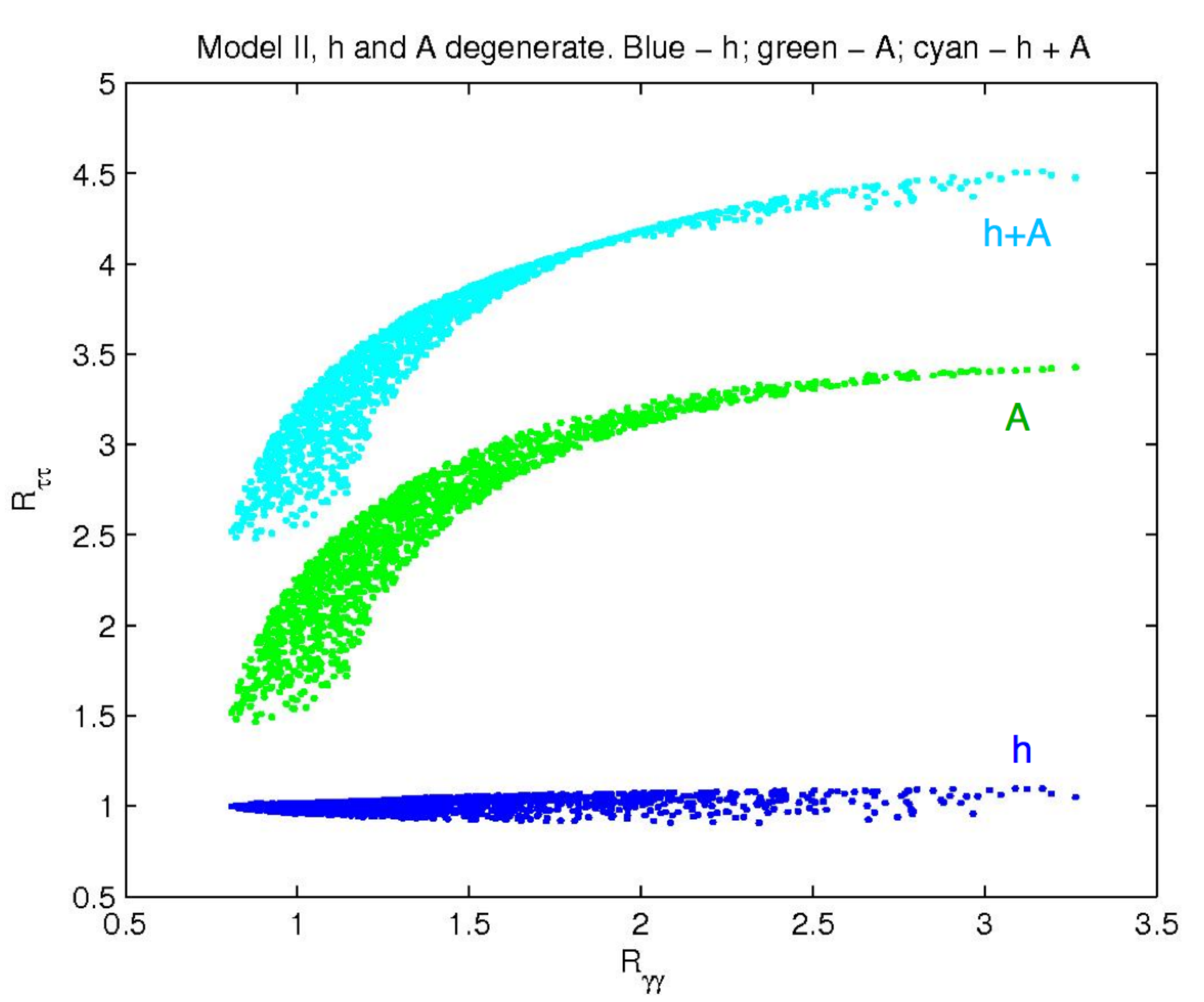}
\caption{Left panel: Allowed region in the
$R^{\textrm{VBF}}_{\gamma \gamma} - R_{\gamma \gamma}$.
Right panel: $R_{\tau \tau}$ as a function of $R_{\gamma \gamma}$ for
$h$ (blue/black),
$A$ (green/gray),
and the total observable rate,
obtained by summing the rates with intermediate $h$ and $A$ (cyan/light-gray).}
\label{fig:vbf_vs_totII}
\end{figure}
The left panel of Fig.~\ref{fig:vbf_vs_totII} exhibits
the allowed region in the
$R^{\textrm{VBF}}_{\gamma \gamma}$--$R_{\gamma \gamma}$
plane,
for our set of points.
In stark contrast with the Type-I 2HDM,
here $R^{\textrm{VBF}}_{\gamma \gamma}$ is predicted
to lie in a rather narrow region close to $0.9$.
This is due to the much narrower range obtained
for $\sin{(\beta - \alpha)}$,
which is constrained to be near 1 in the Type-II scenario,
and implies that $\cos\alpha\simeq \sin\beta$
and $\sin\alpha\simeq -\cos\beta$ (i.e.~the decoupling limit).
Consequently, the Type-II couplings of $h$ to fermion pairs are close to their SM values.
It immediately follows that $R^{\textrm{VBF}}_{\gamma \gamma}$
should be close to 1 as indicated by the left panel of
in Fig.~\ref{fig:vbf_vs_totII}.
If the experimental data were to indicate a similar and sizeable enhancement
of both $R^{\textrm{VBF}}_{\gamma \gamma}$ and $R_{\gamma \gamma}$,
then the $m_h \simeq m_A$ Type-II scenario would be excluded.

As for $R_{\tau \tau}$,
the largest allowed values are less extreme than those of the
Type-I model,
as can be seen from the right panel of Fig.~\ref{fig:vbf_vs_totII}.
%
%
The lower limit on $R_{\tau \tau}$
depends on the value of $R_{\gamma \gamma}$.
For example,
for
$R_{\gamma \gamma} \sim 1.5$,
the total $R_{\tau \tau}$  cannot be
smaller than about $3.5$.
Notice that the right panel of
Fig.~\ref{fig:vbf_vs_totII} shows
a rather strong correlation
between $R_{\tau \tau}$ and $R_{\gamma \gamma}$ in the Type-II model,
while the Type-I model points exhibited in the left panel of
Fig.~\ref{fig:phph_vs_bbI}
are much more dispersed
in the $ R_{\gamma \gamma}$--$R_{\tau \tau}$ plane.

\section{Degenerate $h$ and $H$}

We now turn to the possibility that
$m_h \simeq m_H$.
We have generated sets of parameters such that
$m_A$ and $m_{H^\pm}$ lie above 500 GeV.
For the case in which $h$ and $H$ are nearly degenerate in mass,
we have found that the $S$, $T$, and $U$ constraints
force $m_{H^\pm} \sim m_A$,
to within 10\%.
In this case,
an enhancement of the $\gamma\gamma$ signal rate is visible (although not as pronounced)
even for values of $\tan{\beta}$ somewhat larger than 2.
As a result, we have focused our scan in a $\tan\beta$
regime between $0.5$ and $5$.

\subsection{Type-I 2HDM}

We begin with the Type-I 2HDM.
In the case of a mass degenerate $h$, $A$ pair discussed previously
$\sin{(\beta - \alpha)} > 0.85$ in both the Type-I and Type-II scenarios,
with only a mild correlation with $\tan{\beta}$.
In contrast, employing the same simplified scenario that was used
in obtaining \eq{RZZ_I_simplified},
we now obtain
\begin{equation}
R_{ZZ}
=
R^h_{ZZ} + R^H_{ZZ}
=
\frac{\cos^2{\alpha}}{\sin^2{\beta}}
\sin^2{(\beta-\alpha)} \frac{\sin^2{\beta}}{\cos^2{\alpha}}
+
\frac{\sin^2{\alpha}}{\sin^2{\beta}}
\cos^2{(\beta-\alpha)} \frac{\sin^2{\beta}}{\sin^2{\alpha}}
= 1,
\label{RZZ_III_simplified}
\end{equation}
where we have used the couplings of Table~\ref{tab:couplings}.
In fact,
even using all production mechanisms,
we find that $R_{ZZ} \sim 1$.  Consequently,
\eq{rzz} places almost no constraint on the value of
$\sin{(\beta - \alpha)}$.

\begin{figure}[t!]
\centering
\includegraphics[width=3.5in,angle=0]{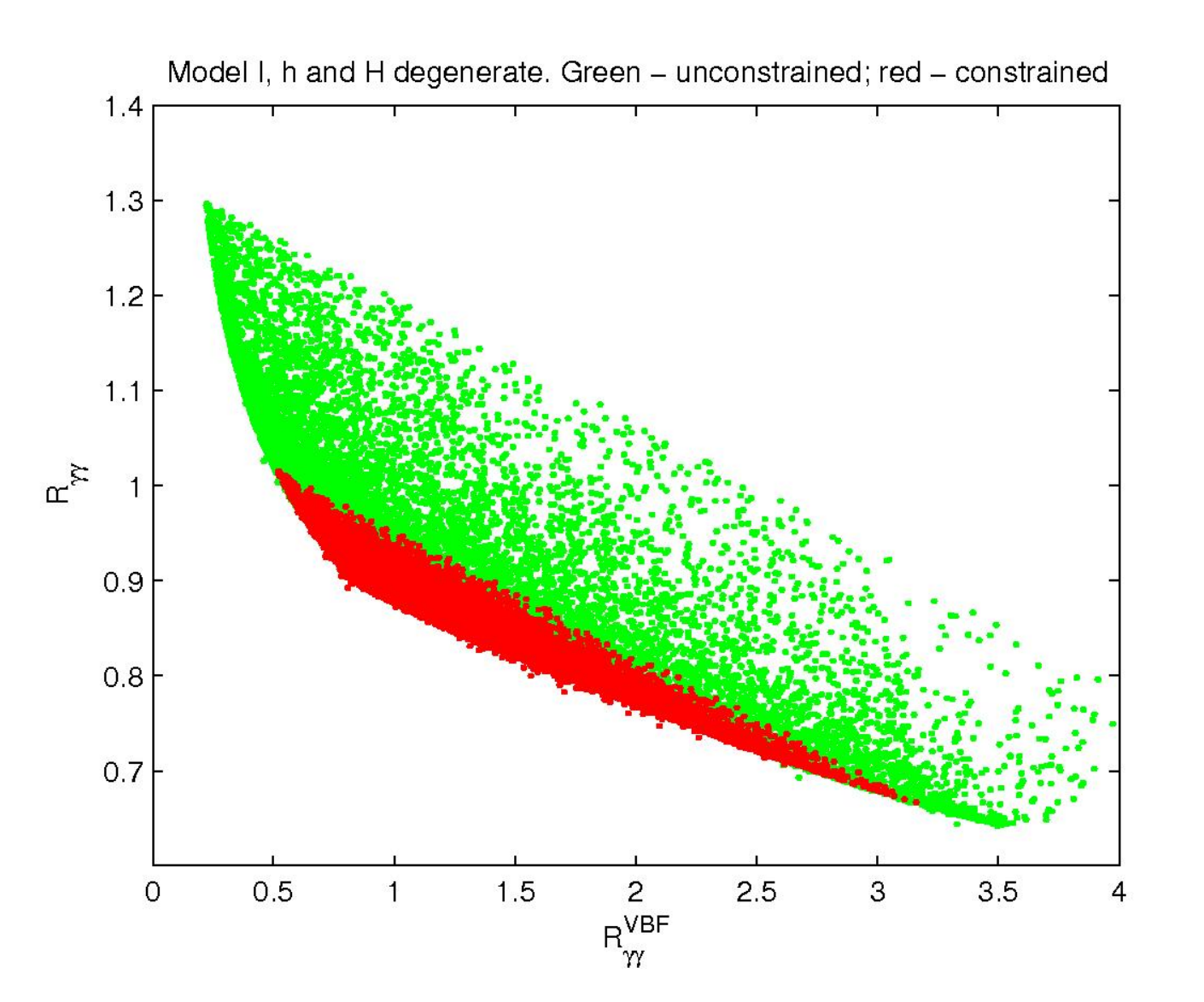}
\hspace{-.3cm}
\includegraphics[width=3.5in,angle=0]{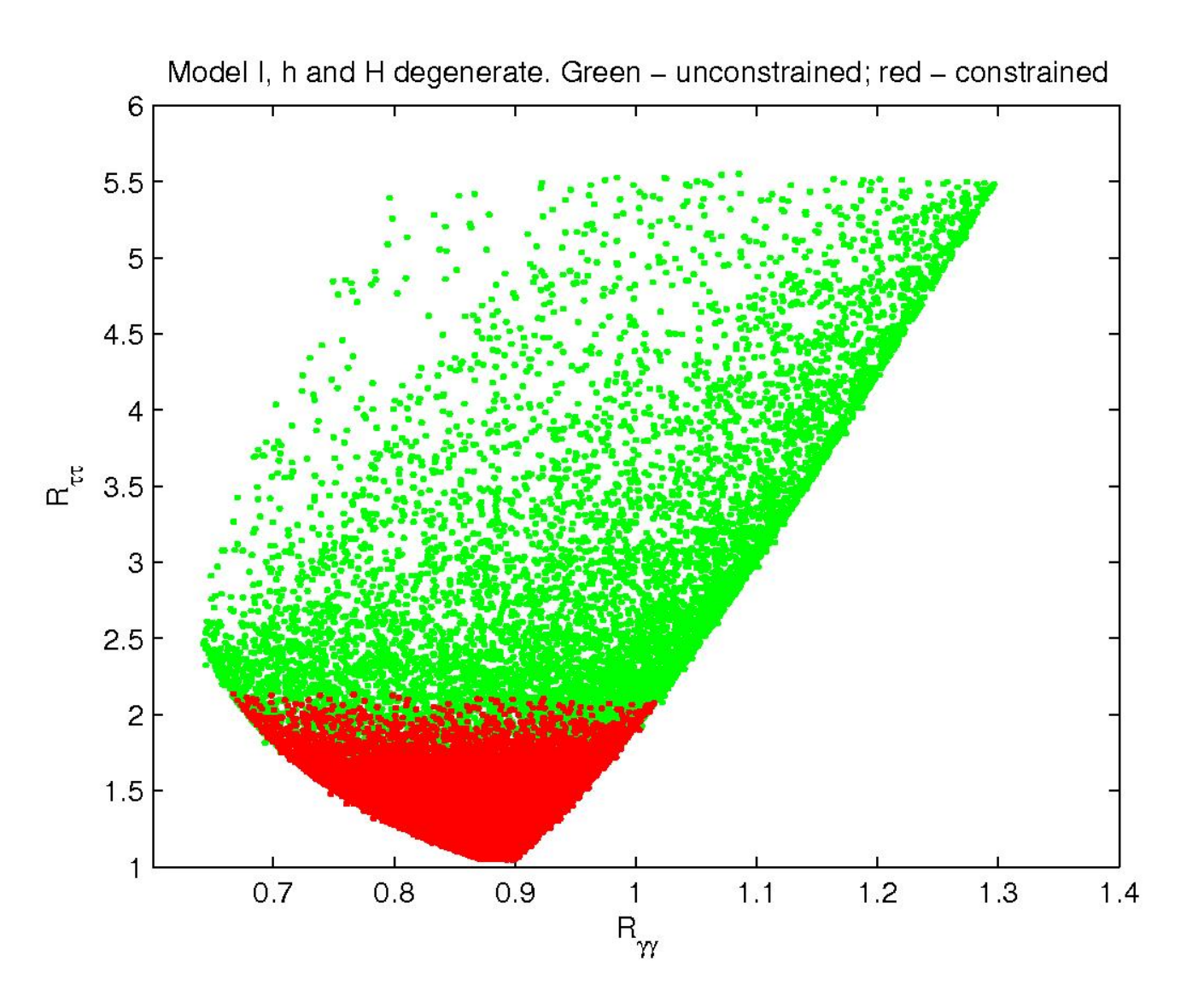}
\caption{Left panel: Allowed region in the
$R^{\textrm{VBF}}_{\gamma \gamma} - R_{\gamma \gamma}$
plane for the constrained (red/black) and unconstrained (green/gray)
scenarios. Right panel: Allowed region in the
$R_{\gamma \gamma} - R_{\tau \tau}$ plane for the
constrained (red/black) and unconstrained (green/gray) scenarios.}
\label{fig:vbf_vs_totI_hH}
\end{figure}

The left panel of Fig.~\ref{fig:vbf_vs_totI_hH}
exhibits the allowed region in the
$R^{\textrm{VBF}}_{\gamma \gamma}$--$R_{\gamma \gamma}$
plane, for the constrained (red/black) and unconstrained
(green/gray) scenarios.
The total $R_{\gamma \gamma}$ cannot differ significantly from
its SM value.
In particular, the maximal allowed $\gamma\gamma$ enhancement
is about $1.3$ for the unconstrained
scenario, and none of the enhanced values survive after
imposing the constraints from $B$ physics.
In contrast,
$R^{\textrm{VBF}}_{\gamma \gamma}$ can be either very close
to vanishing
or take values as large as 4 in the unconstrained scenario and
as large as 3 in the constrained scenario.

The possibility of an enhanced $R^{\textrm{VBF}}_{\gamma \gamma}$ arises
due to the fact that $\alpha$ is not especially
constrained by the requirement of \eq{rzz}.
In particular, the limit of $\sin\alpha=\pm 1$ [$\sin\alpha=0$]
corresponds to the fermiophobic limit for $h$
[$H$].  In this limit, the dominant fermiophobic Higgs boson decay
channels are $WW^*$ and $ZZ^*$, in which
case we expect the corresponding branching ratio into
$\gamma\gamma$ (which should be approximately given by the ratio of
the $\gamma\gamma$ and $WW^*$ partial widths) to be enhanced by a factor of roughly 5
relative to its value in the SM.
Thus,
even though the VBF production cross-section is roughly given by
its SM value,
it is not surprising that one can achieve values
of $R^{\textrm{VBF}}_{\gamma \gamma}$ as large
as shown in the left panel of Fig.~\ref{fig:vbf_vs_totI_hH}.
Moreover, one also expects a reduced value of
$R_{\gamma \gamma}$, since in the fermiophobic limit
the size of the gluon-gluon fusion cross section
(which depends on the coupling of the Higgs boson
to $t\bar{t}$) is significantly suppressed.
Assuming that the VBF production mechanism now dominates,
it follows that the Higgs production cross-section
has been reduced by roughly a factor of 10 relative to
the SM Higgs production cross-section via gluon-gluon
fusion.  This reduction factor of 10 cannot be completely
compensated by an increase in the Higgs to $\gamma\gamma$ branching
ratio, which implies that $R_{\gamma \gamma}<1$ in
the fermiophobic regime,
as shown in the left panel of Fig.~\ref{fig:vbf_vs_totI_hH}.

The right panel of
Fig.~\ref{fig:vbf_vs_totI_hH} exhibits
$R_{\tau \tau}$ as a function of $R_{\gamma \gamma}$.
It is interesting to compare the Type-I scenarios exhibited by the left panel of
Fig.~\ref{fig:phph_vs_bbI},
which holds for the case of a mass-degenerate $h$, $A$ pair,
with the right panel of Fig.~\ref{fig:vbf_vs_totI_hH},
which holds for the case of a mass-degenerate $h$, $H$ pair.
In contrast to the former scenario, where $R_{\gamma \gamma}$ could reach
about $2$ while $R_{\tau \tau}$ could reach $11$, in the present scenario
$R_{\gamma \gamma}$ can be at most $1.3$
while $R_{\tau \tau}$ is smaller than $5.5$.
Nevertheless,
the lower bound on $R_{\tau \tau}$ for
$R_{\gamma \gamma} > 1.1$ is more
stringent in the latter scenario than in the former.
Indeed,
when $m_h \simeq m_A$ and $R_{\gamma \gamma} = 1.2$,
we predict $R_{\tau \tau} > 2$,
whereas
when $m_h \simeq m_H$ and $R_{\gamma \gamma} = 1.2$,
we predict $R_{\tau \tau} > 3.5$.
Again, in the constrained scenario, $R_{\gamma \gamma} $
has to be below $1$ while $R_{\tau \tau}$ cannot be above $2$.
The constrained scenario makes a very strong prediction:
$R_{\gamma \gamma} $ has to be SM-like or smaller, whereas
$R_{\tau \tau}$ must be above the SM prediction.

\subsection{Type-II 2HDM}

\begin{figure}[hb!]
\centering
\includegraphics[width=3.5in,angle=0]{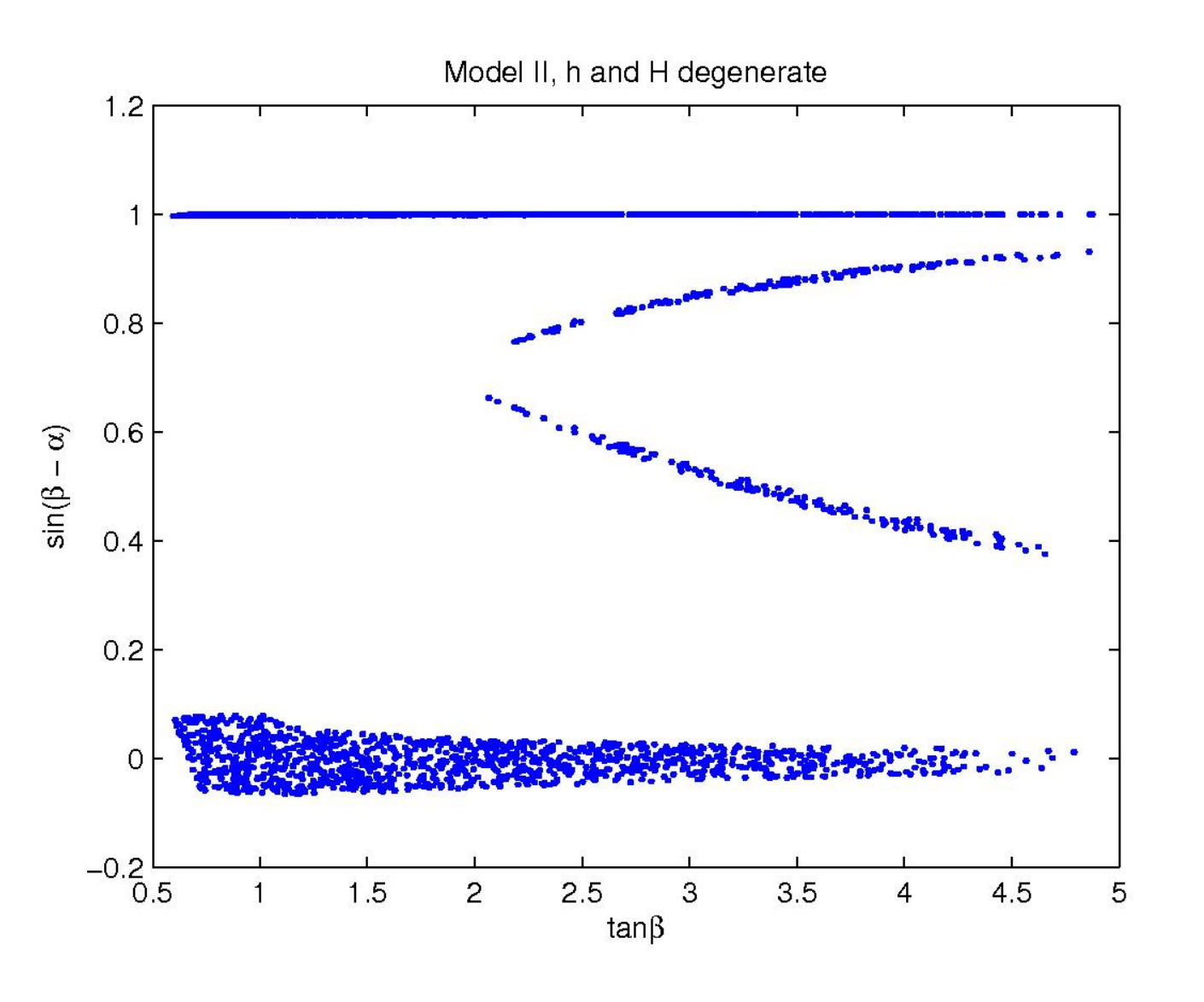}
\hspace{-.3cm}
\includegraphics[width=3.5in,angle=0]{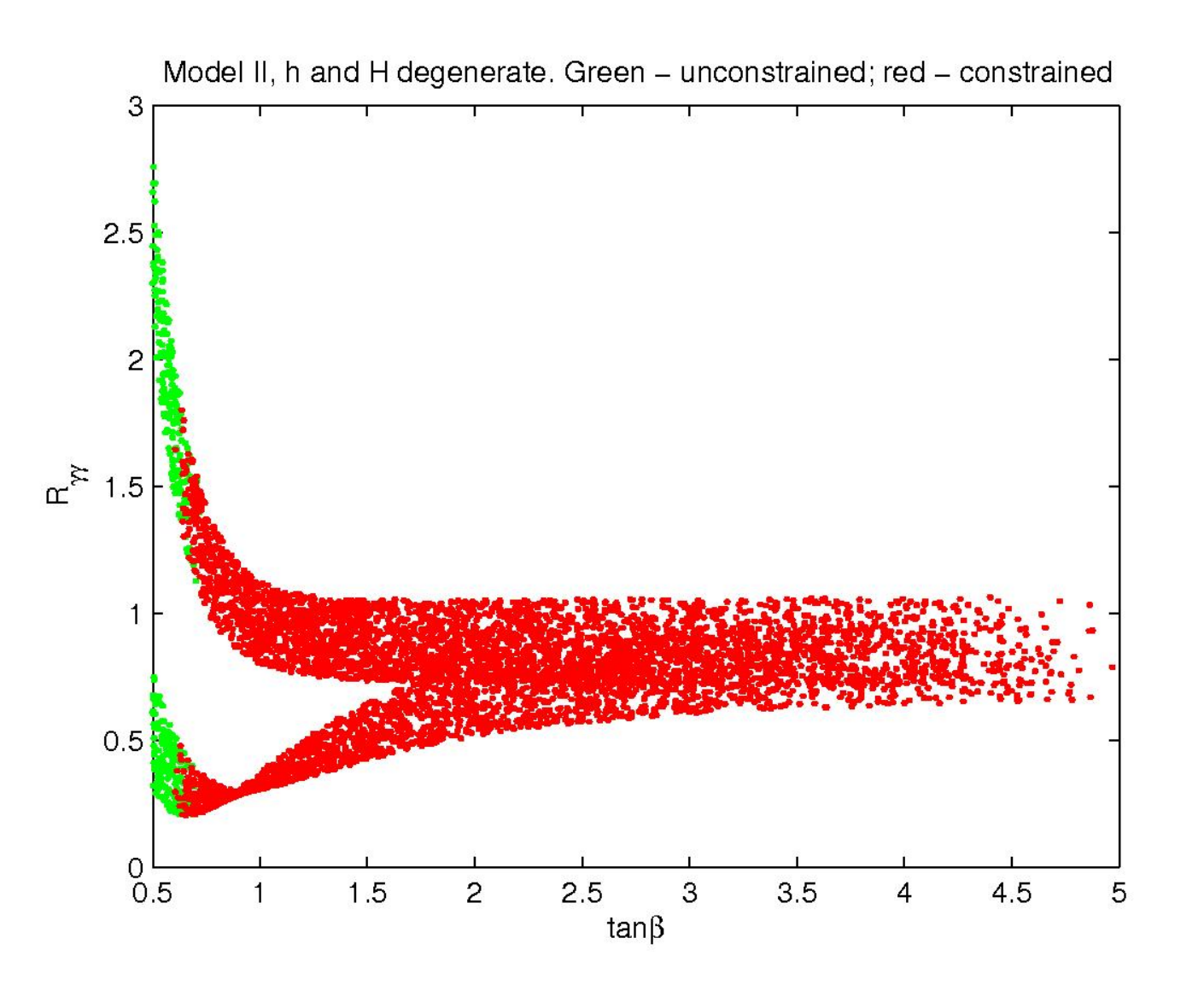}
\caption{Left panel: Values obtained in the
$ \tan{\beta}- \sin{(\beta - \alpha)}$ plane for the points generated,
which satisfy $0.8 < R_{\gamma \gamma} < 1.5$.
Right panel: Values for $R_{\gamma \gamma}$ as a function of
$ \tan{\beta}$ for the constrained (red/black)
and unconstrained (green/gray) scenarios.}
\label{fig:tanb_sin_hHII_NEW2}
\end{figure}

In this case there is a strong correlation between
$\sin{(\beta - \alpha)}$ and $\tan{\beta}$.
Employing the same simplified scenario that was used
in obtaining \eq{RZZ_I_simplified}
and using the couplings of Table~\ref{tab:couplings}, we now obtain
\beq
R_{ZZ}
=
R^h_{ZZ} + R^H_{ZZ}=
\frac{\cos^2{\alpha}}{\sin^2{\beta}}
\sin^2{(\beta-\alpha)} \frac{\cos^2{\beta}}{\sin^2{\alpha}}
+
\frac{\sin^2{\alpha}}{\sin^2{\beta}}
\cos^2{(\beta-\alpha)} \frac{\cos^2{\beta}}{\cos^2{\alpha}}\,.
\eeq
The above result can be expressed directly in terms of $\tan\beta$ and $\sin(\beta-\alpha)$ by
employing \eqs{sa}{ca},
\beq
R_{ZZ}= \frac{[\tan^2\beta+\cot^2\beta+2]c^2_{\beta-\alpha}s^2_{\beta-\alpha}
+[\tan\beta(c^2_{\beta-\alpha}-s^2_{\beta-\alpha})-2c_{\beta-\alpha}s_{\beta-\alpha}]^2}
{[\tan\beta(c^2_{\beta-\alpha}-s^2_{\beta-\alpha})-(1-\tan^2\beta)c_{\beta-\alpha}s_{\beta-\alpha}]^2}\,,
\label{RZZ_IV_simplified}
\eeq
where $c_{\beta-\alpha}\equiv\cos(\beta-\alpha)$ and $s_{\beta-\alpha}\equiv\sin(\beta-\alpha)$.
Thus, \eqs{rzz}{RZZ_IV_simplified}
impose a complicated constraint in the
$\sin(\beta-\alpha)$--$\tan\beta$ plane.
For example, for model points where
$0.8 < R_{\gamma \gamma} < 1.5$,
the allowed regions in the $\sin(\beta-\alpha)$--$\tan\beta$
plane are shown in the left panel of
Fig.~\ref{fig:tanb_sin_hHII_NEW2}.
The uppermost (lowermost) two bands correspond to
a dominant $h$ ($H$) contribution to $R_{\gamma \gamma}$.
In the band around $\sin{(\beta - \alpha)} \sim 1$,
$h$ is gaugephilic and $H$ is gaugephobic.
In the second band below,
the contribution of $h$ to $R_{\gamma \gamma}$ is still dominant,
with a small contribution by $H$.
In the next band,
the roles of $h$ and $H$ are reversed,
and in the band around $\sin{(\beta - \alpha)} \sim 0$,
$H$ is gaugephilic while $h$ is gaugephobic.
\begin{figure}[t!]
\centering
\includegraphics[width=3.5in,angle=0]{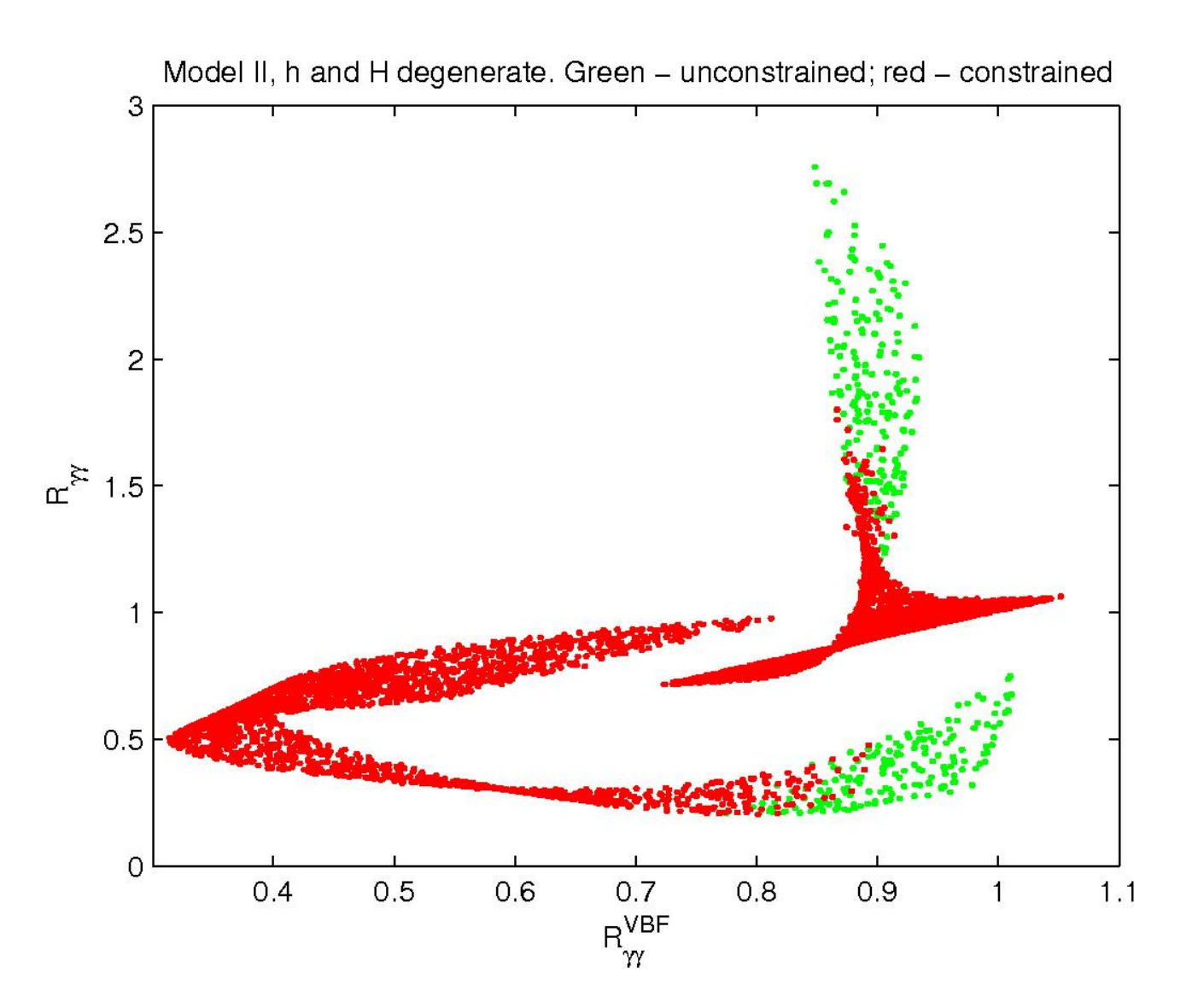}
\hspace{-.3cm}
\includegraphics[width=3.5in,angle=0]{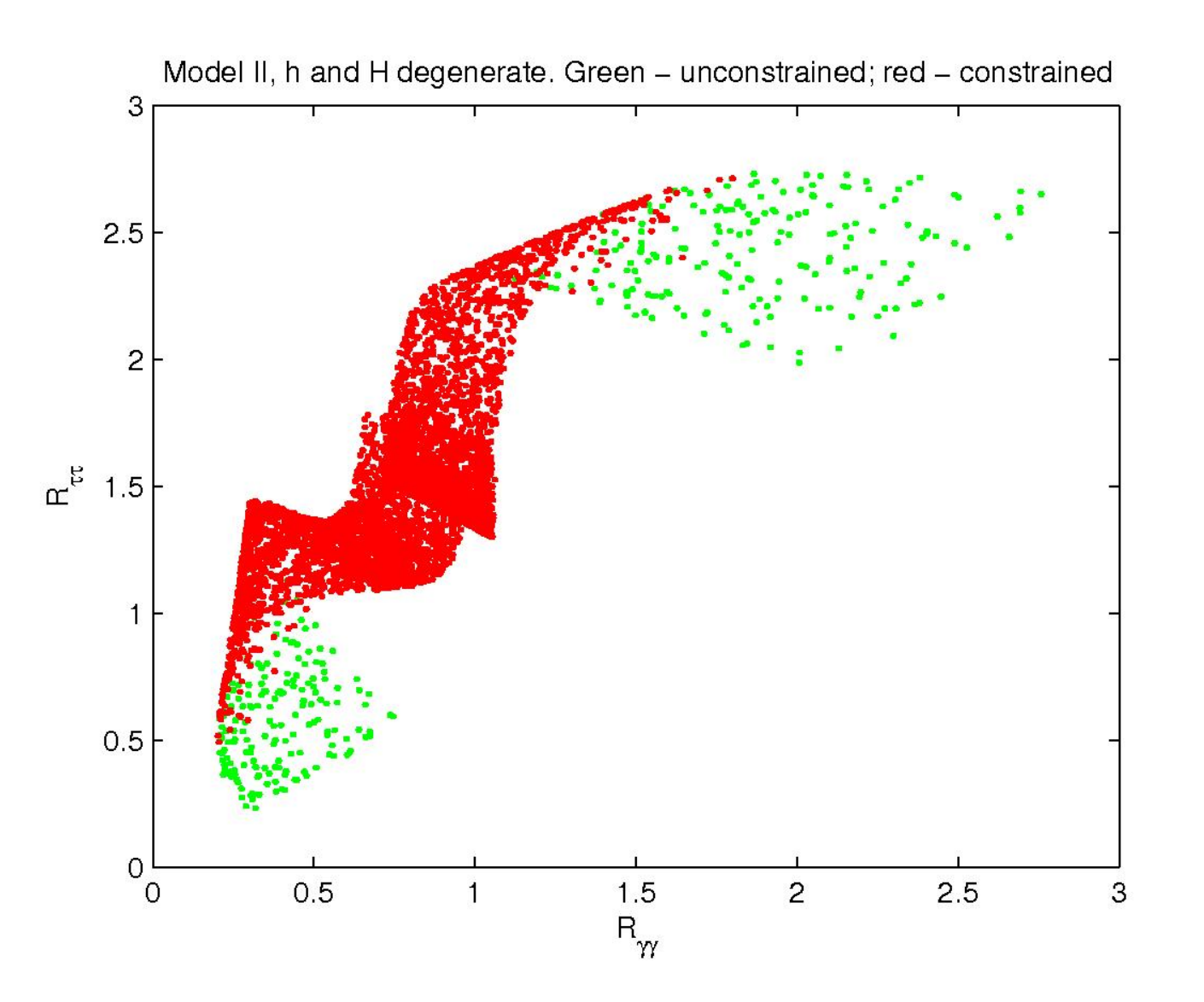}
\caption{Left panel: Allowed region in the
$R^{\textrm{VBF}}_{\gamma \gamma}$--$R_{\gamma \gamma}$
plane for the constrained (red/black) and unconstrained
(green/gray) scenarios. Right panel: Allowed region in the
$R_{\gamma \gamma}$--$R_{\tau \tau}$ plane for the constrained
(red/black) and unconstrained (green/gray) scenarios.}
\label{fig:vbf_vs_totII_hH}
\end{figure}
The band structure exhibited in the left panel of Fig.~\ref{fig:tanb_sin_hHII_NEW2}
impacts the
$R_{\gamma \gamma}$ dependence on $\tan{\beta}$,
as shown in the right panel of Fig.~\ref{fig:tanb_sin_hHII_NEW2}.
The generated set of points are
plotted in the
$R^{\textrm{VBF}}_{\gamma \gamma}$--$ R_{\gamma \gamma}$
plane in the left panel of
Fig.~\ref{fig:vbf_vs_totII_hH}, whereas the allowed regions
in the $R_{\gamma \gamma}$--$R_{\tau \tau}$ plane
is shown in  the right panel of Fig.~\ref{fig:vbf_vs_totII_hH}
for the constrained (red/black) and unconstrained (green/gray) scenarios.
It is instructive to compare the $m_h \simeq m_H$ type-II model analyzed
in this section with
the $m_h \simeq m_A$ type-II model examined in Section 4.2.
Comparing left panels of Fig.~\ref{fig:vbf_vs_totII} and
Fig.~\ref{fig:vbf_vs_totII_hH},
we see that the former scenario allows for larger values
of $R_{\gamma \gamma}$.
In contrast,
the values of $R^{\textrm{VBF}}_{\gamma \gamma}$
are very constrained in the former scenario,
while in the latter scenario these values can range from $0$ to $1.1$.
Comparing the right panels of Fig.~\ref{fig:vbf_vs_totII}
and Fig.~\ref{fig:vbf_vs_totII_hH},
we see that $2.5 < R_{\tau \tau} < 4.5$ in the case of $m_h=m_A$,
whereas $0.2 < R_{\tau \tau} < 2.7$ in the case of $m_h=m_H$.

Once the constraints of $B$ physics are applied,
the most striking difference from the results presented above
is that the value of $R_{\gamma \gamma}$ in the constrained scenario must be below $1.5$.
The allowed ranges of $R^{\textrm{VBF}}_{\gamma \gamma}$
and $R_{\tau \tau}$ are quite narrow.
However,
a definite
measurement of $R_{\gamma \gamma} = 1.5$ would force
$R_{\tau \tau} \approx  2.5$.
It is also interesting to compare the
$m_h \simeq m_H$ Type-II model analyzed here
with the $m_h \simeq m_H$ Type-I model examined in Section 5.1.
Here the enhancement in $R_{\gamma \gamma}$ can be larger, whereas
the upper bounds on
the enhancements in $R^{\textrm{VBF}}_{\gamma \gamma}$
and $R_{\tau \tau}$ are somewhat reduced.
In summary,
it is more difficult to generate an enhancement of $R_{\gamma \gamma}$
in the $m_h \simeq m_H$ case as compared to the $m_h \simeq m_A$ case.
On the other hand, the corresponding enhancement of
$R_{\tau \tau}$ is also reduced.

\section{Degenerate $H$ and $A$}

Now we turn to the case where $m_h < m_H \simeq m_A  \simeq $ 125 GeV.
In the previous cases we have studied,
we could keep $m_{H^\pm} > 500\, \textrm{GeV}$.
In this case,
there are no points with $m_{H^\pm} > 200\, \textrm{GeV}$
that survive the constraints coming from
boundedness from below,
unitarity,
and precision electroweak measurements.
Since the constraints from $b \to s \gamma$ imply
$m_{H^\pm} > 360\, \textrm{GeV}$ in the Type-II 2HDM,
the whole of parameter space with $m_H \simeq m_A \simeq$ 125 GeV
is excluded in the Type-II model.
Thus we examine the $m_A \simeq m_H$ scenario in Type-I 2HDM,
keeping $0.5 < \tan{\beta} < 5$.

\begin{figure}[t!]
\epsfysize=7cm
\centerline{\epsfbox{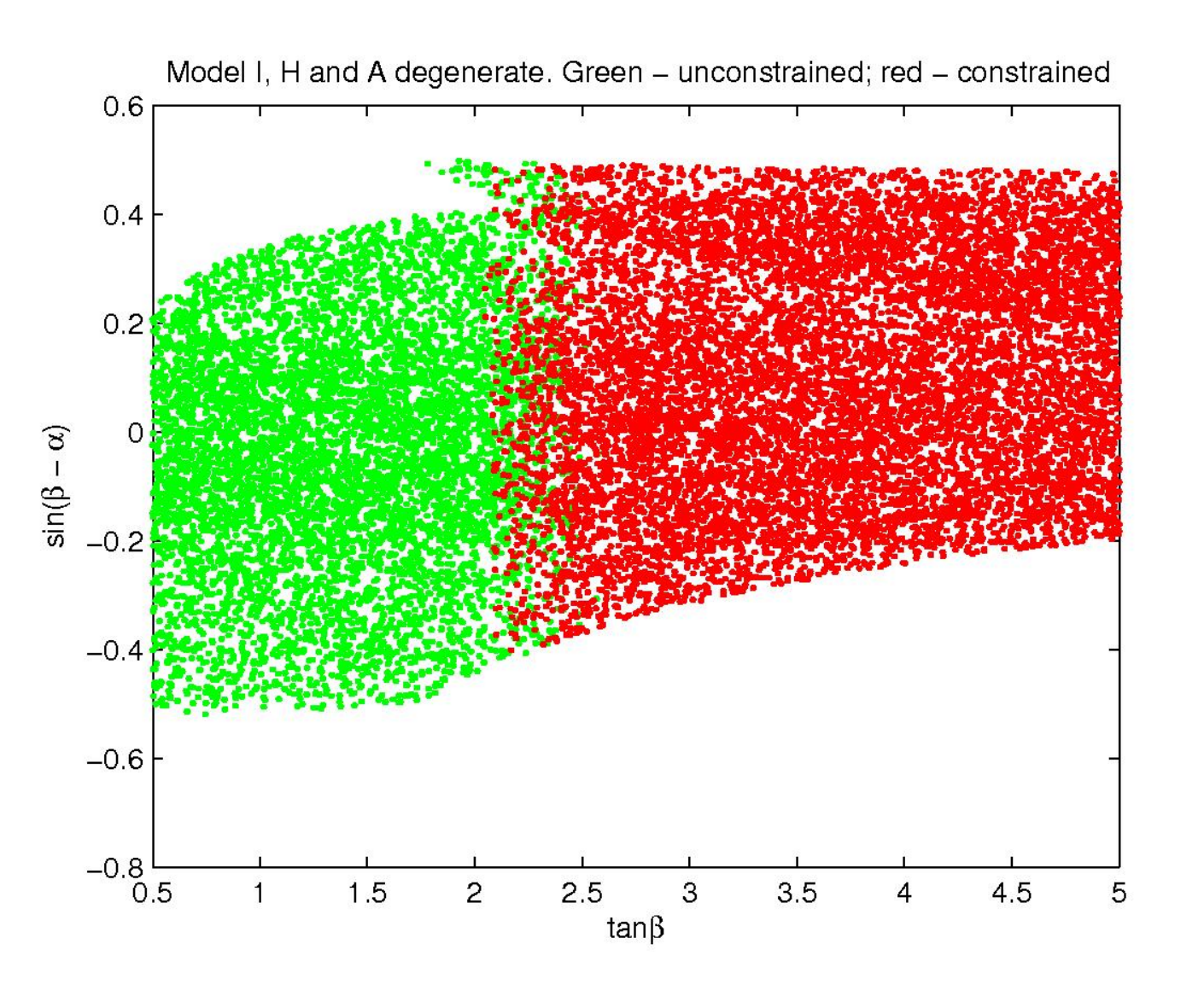}}
\caption{Values obtained in the
$ \tan{\beta}$--$\sin{(\beta - \alpha)}$ plane for the
points generated for the constrained (red/black)
and unconstrained (green/gray) scenarios.
}
\label{fig:tanb_sin_HhAI}
\end{figure}

\begin{figure}[h!]
\centering
\includegraphics[width=3.5in,angle=0]{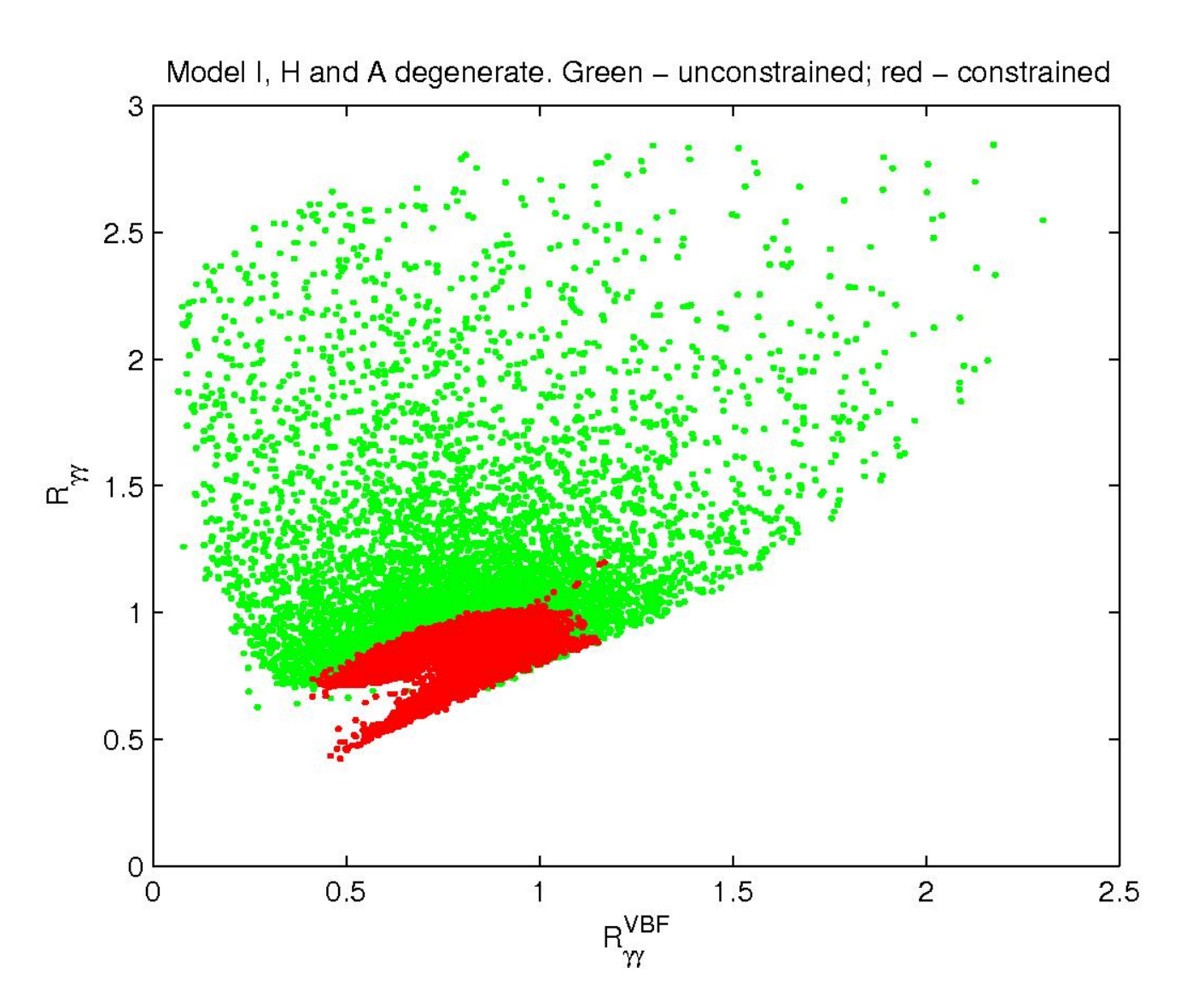}
\hspace{-.3cm}
\includegraphics[width=3.5in,angle=0]{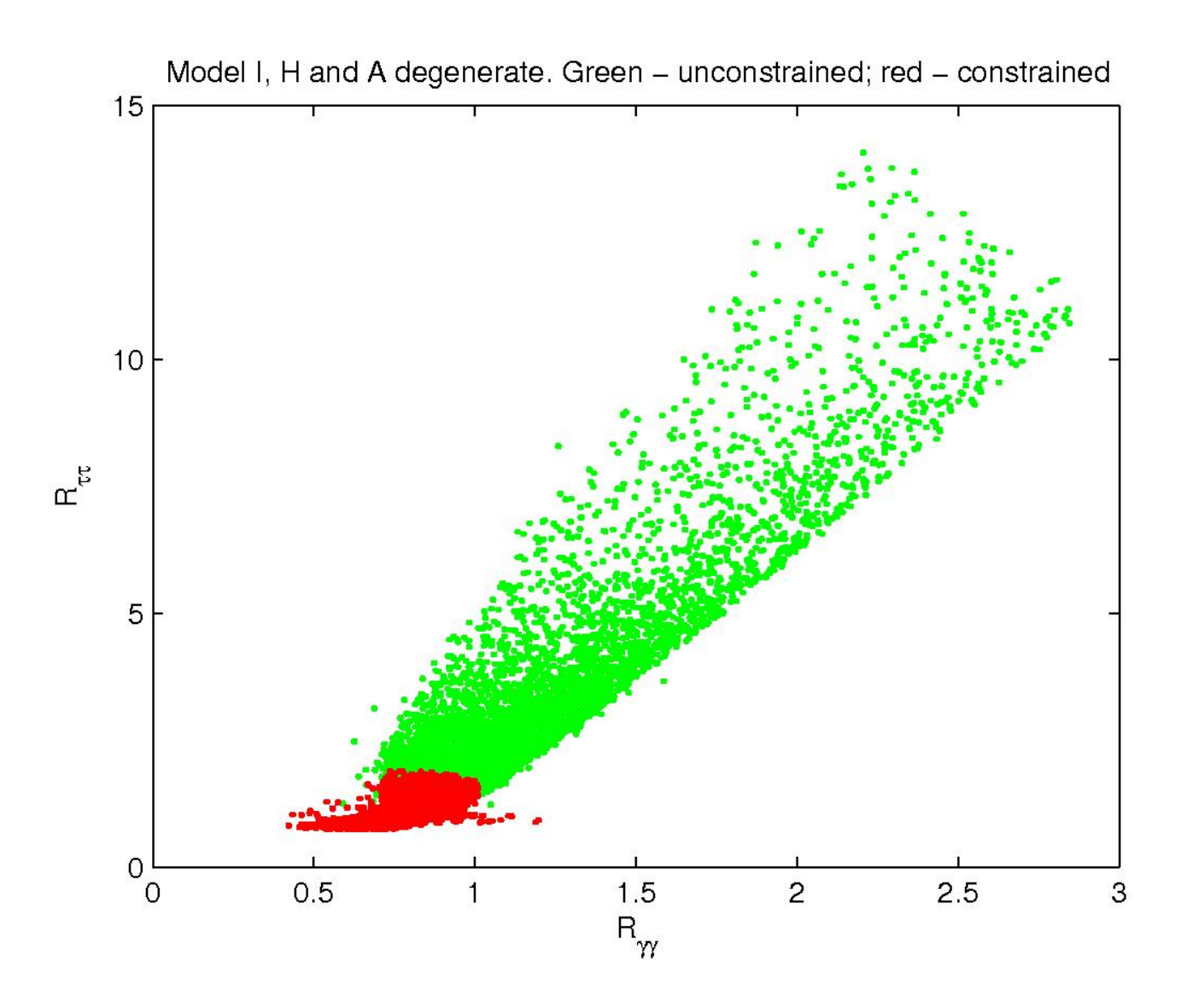}
\caption{Left panel: Allowed region in the
$R^{\textrm{VBF}}_{\gamma \gamma}$--$R_{\gamma \gamma}$
plane for the constrained (red/black) and unconstrained
(green/gray) scenarios. Right panel: Allowed region in the
$R_{\gamma \gamma}$--$R_{\tau \tau}$ plane for the constrained
(red/black) and unconstrained (green/gray) scenarios.}
\label{fig:vbf_vs_totI_AHh}
\end{figure}

If $m_h < m_H/2$,
then the branching ratio for $H \to h h$ would be
close to one,
greatly suppressing the $Z Z^*$ signal
(to which $A$ cannot contribute).
As a result, we shall assume that $m_h > m_H/2$.
In particular, we examine the mass range
$65\, \textrm{GeV} < m_h < 110 \, \textrm{GeV}$.
Such a low-mass $h$ could have been produced at LEP
via the $ZZh$ vertex,
which is suppressed by a factor of $\sin{(\beta - \alpha)}$
in the 2HDM.  The model points surviving the LEP constraints~\cite{mssmhiggs},
in which \eq{rzz} is satisfied due to $H$ production and decay,
are shown in
the $ \tan{\beta}$--$\sin{(\beta - \alpha)}$ plane of
Fig.~\ref{fig:tanb_sin_HhAI} for the constrained (red/black)
and unconstrained (green/gray) scenarios.
Note that because the charged Higgs boson in now forced to be light,
the $B$-physics constraints imply a value of $\tan \beta$ above $2$.
As expected,
the points are centered around
$\sin{(\beta - \alpha)} = 0$,
but values as large as $|\sin{(\beta - \alpha)}| \sim 0.5$
are possible.

The generated set of points are plotted in the
$R^{\textrm{VBF}}_{\gamma \gamma}$--$R_{\gamma \gamma}$
plane in the left panel of
Fig.~\ref{fig:vbf_vs_totI_AHh} for the constrained
(red/black) and unconstrained (green/gray) scenarios.
Contrary to the previous scenarios with $m_h \simeq m_A$ or $m_h \simeq m_H$
in the Type-I model,
both $R^{\textrm{VBF}}_{\gamma \gamma}$ and $R_{\gamma \gamma}$ now
have a much wider range of variation and can reach $2.3$ simultaneously.
However,
in the constrained scenario the $R^{\textrm{VBF}}_{\gamma \gamma}$
values drop
to a maximum of around $1.2$ while $R_{\gamma \gamma}$ drops
to a maximum close to $1$.
The allowed region in the $R_{\gamma \gamma}$--$R_{\tau \tau}$ plane
for this set of points
is shown in the right panel of Fig.~\ref{fig:vbf_vs_totI_AHh}.
We note that the constrained scenario lives
in a region very close to the SM prediction.

\section{Degenerate $h$, $H$ and $A$}

In this section we discuss the case where $m_h \simeq m_H \simeq m_A \simeq 125$ GeV, i.e.~all three neutral scalars
are nearly mass-degenerate.  As in the previous scenario with
$m_H \simeq m_A $,
the electroweak precision constraints force the charged Higgs to be light
(below approximately 200 GeV).
Hence, due to the $b \to s \gamma$ bound on the charged Higgs mass, this
scenario is  ruled out in the Type-II 2HDM.

\begin{figure}[t!]
\centering
\includegraphics[width=3.5in,angle=0]{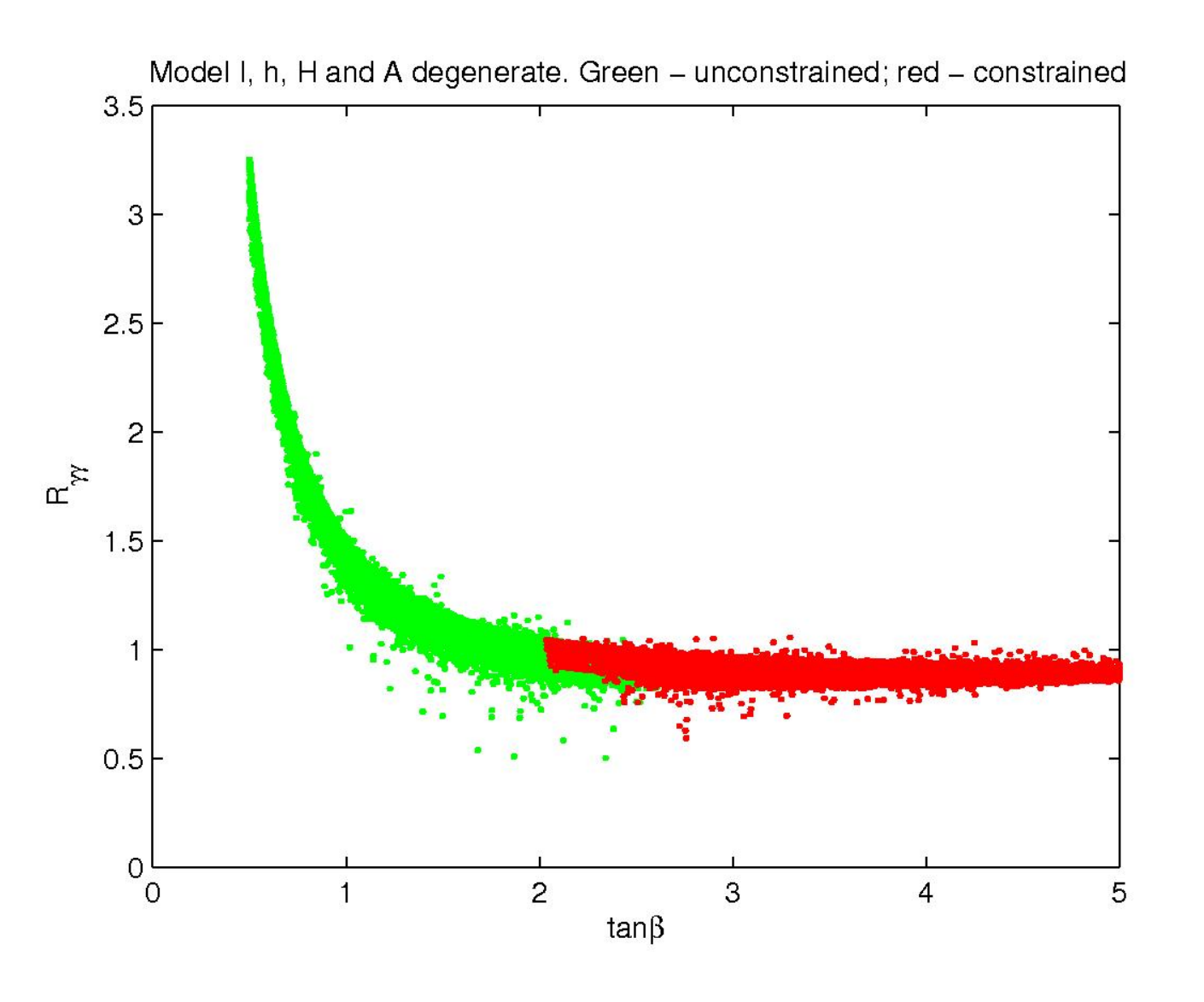}
\hspace{-.3cm}
\includegraphics[width=3.5in,angle=0]{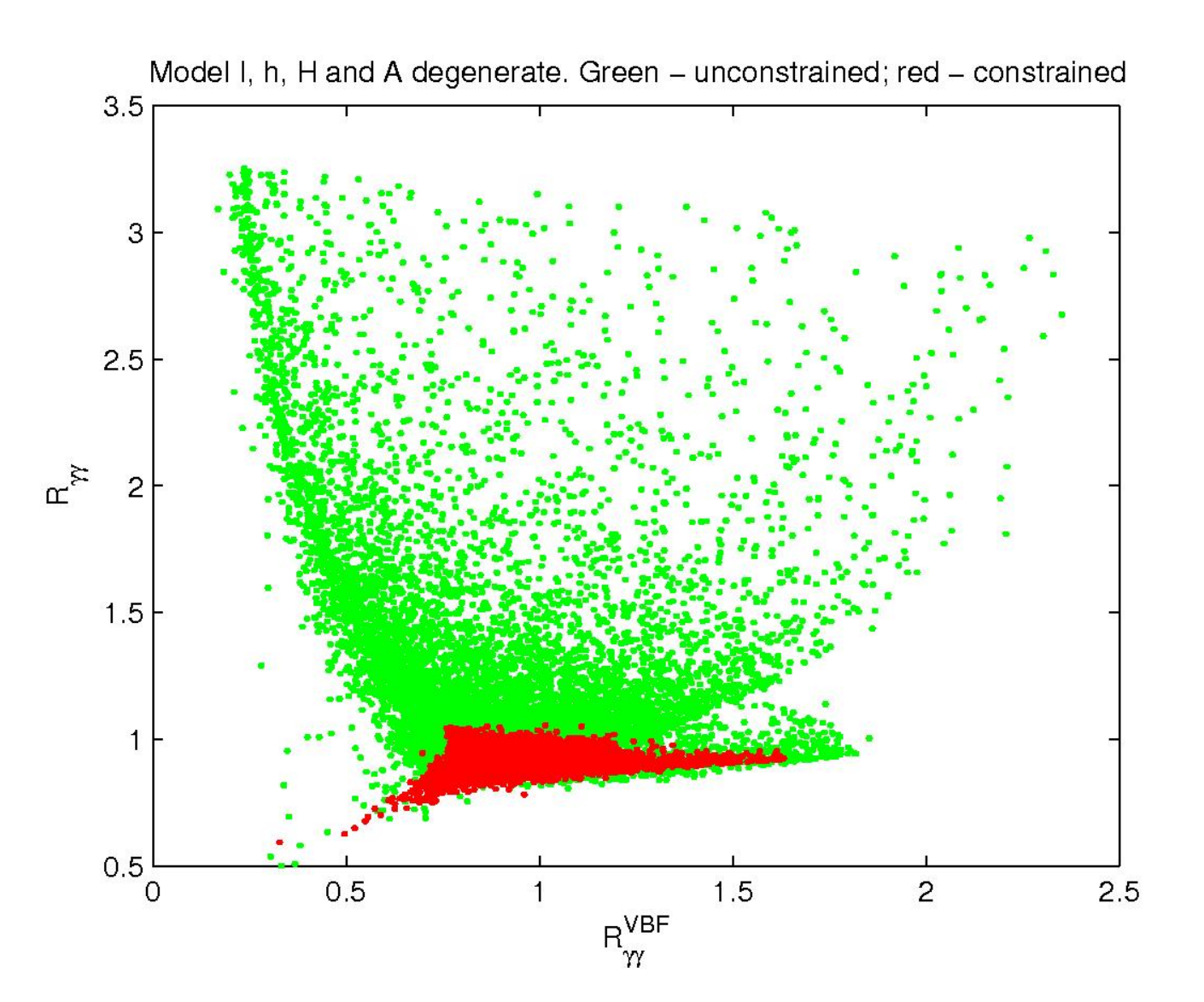}
\caption{Left panel: Total $R_{\gamma \gamma}$
($h$,  $H$ and $A$ summed) as a
function of $\tan \beta$ for the constrained (red/black)
and unconstrained (green/gray)
scenarios.
Right panel: Allowed region in the
$R^{\textrm{VBF}}_{\gamma \gamma} - R_{\gamma \gamma}$
plane for the constrained (red/black) and unconstrained (green/gray) scenarios.}
\label{fig:hAH}
\end{figure}

In the left panel of Fig.~\ref{fig:hAH} we present the total $R_{\gamma \gamma}$, with
$h$,  $H$ and $A$ summed,
as a function of $\tan \beta$ for the constrained (red/black)
and unconstrained
(green/gray) scenarios in the Type-I 2HDM.
This scenario does not differ much from the previous one where again only
Type-I was allowed.
In particular, when all constraints
are imposed, the maximum value of $R_{\gamma \gamma}$ has to
lie very close to $1$.
In the right panel of Fig.~\ref{fig:hAH} we present the allowed region in the
$R^{\textrm{VBF}}_{\gamma \gamma}$--$R_{\gamma \gamma}$
plane for the constrained (red/black) and unconstrained (green/gray) scenarios.
If we now compare this scenario with the three previous ones we conclude
that the differences are not that striking,
especially in the constrained case:
$R^{\textrm{VBF}}_{\gamma \gamma}$ can be
larger
than the SM value reaching 1.7 while $R_{\gamma \gamma}$
has to be very close to the value predicted by the SM.
However, a large value of
$R^{\textrm{VBF}}_{\gamma \gamma}= 1.7$ and $R_{\gamma \gamma} = 1$
is only allowed in the present scenario and in the Type-I 2HDM with $m_H \simeq m_h$.
In fact, the difference between these two scenarios is primarily due to the
theoretical
and experimental constraints: when $m_H \simeq m_h $ a nearly-fermiophobic scenario
is allowed with values of $R^{\textrm{VBF}}_{\gamma \gamma}$ reaching 3.
In contrast, when all the neutral Higgs masses are degenerate the fermiophobic
limit cannot be attained, and $R^{\textrm{VBF}}_{\gamma \gamma}$  reaches
a maximum value of $1.7$.

\section{Conclusions}

In this paper, we have addressed the possibility that an
enhanced $\gamma\gamma$ signal in
Higgs production at the LHC
(under the assumption of a SM-like $ZZ\to 4~{\rm leptons}$
signal)
can be explained in the context of the 2HDM by
a pair of nearly mass-degenerate neutral
Higgs bosons with a mass around 125~GeV.
To analyze this scenario, we have examined
the softly broken $\mathbb{Z}_2$-symmetric and CP-conserving 2HDM with either
Type-I or Type-II Higgs-fermion Yukawa couplings.
We have scanned the resulting 2HDM parameter spaces,
subject to the constraints
of all known experimental observables (excluding the
recently observed anomaly of
$\overline{B}\to D^{(*)}\tau^-\overline{\nu}_\tau$
by the BaBar collaboration, which has not
yet been confirmed by the Belle collaboration).

We find that,
in the degenerate mass scenario,
it is only possible to produce an enhanced $\gamma\gamma$
signal in a region of parameter
space where $\tan\beta$ is near 1.
This result immediately implies that it is not possible to
realize such a scenario in the MSSM, since the additional
constraints imposed by supersymmetry
combined with the LEP search for the Higgs bosons of the
MSSM rule out values of $\tan\beta$ below about 2~\cite{mssmhiggs}.

In particular, we have demonstrated that an
enhanced $\gamma\gamma$ signal could be due to
a nearly mass-degenerate $h$, $A$ pair or $h$, $H$ pair.
(In the case where $H$ and $A$ are nearly mass-degenerate,
it is much more difficult to
achieve a significant enhancement
in the $\gamma\gamma$ signal subject to all the experimental constraints.)
In the parameter region
corresponding to an enhanced $\gamma\gamma$ signal,
we generically expect an enhanced inclusive $\tau^+\tau^-$
signal, due to the contribution of the second
mass-degenerate Higgs state in addition to the $\tau^+\tau^-$
signal produced by the SM-like Higgs state.

The absence of an enhanced $\tau^+\tau^-$ signal would
be strong evidence against the scenario proposed in this paper.
However, if both a $\gamma\gamma$ and  $\tau^+\tau^-$ enhancement
are confirmed, then one can probe the nature of the Higgs--fermion
Yukawa couplings by distinguishing $\gamma\gamma$ signal events
that arise from vector boson fusion (VBF).  For example, we find that
an enhanced VBF $\gamma\gamma$ rate is possible in Type-I models but
is not possible in Type-II models.

Of course, ultimately the verification of the mass-degenerate scenario
would require the observation of two separate scalar states.  As we do
not expect these states to be exactly mass-degenerate, it is possible
that the inherent experimental mass resolutions of the ATLAS and CMS experiments
could eventually be sensitive to the presence of two (or more)
approximately mass-degenerate scalar states.
In particular, in the mass-degenerate scenarios studied in this paper, the $\gamma\gamma$ signal
is a consequence of the production and decay of both mass-degenerate Higgs
bosons, whereas the $ZZ^*\to 4$~lepton signal derives primarily from the production
and decay of a single Higgs boson with SM-like couplings to vector boson pairs.
Thus, an observation of slightly different invariant masses in the $\gamma\gamma$
and  $ZZ^*\to 4$~lepton channels would be strong evidence that a scenario
similar to the ones examined in this paper is realized in nature.

\acknowledgments{
The works of P.M.F. and R.S. are supported in part by the Portuguese
\textit{Funda\c{c}\~{a}o para a Ci\^{e}ncia e a Tecnologia} (FCT)
under contract PTDC/FIS/117951/2010, by FP7 Reintegration Grant, number PERG08-GA-2010-277025,
and by PEst-OE/FIS/UI0618/2011.
The work of H.E.H. is supported
in part by the U.S. Department of Energy, under grant
number DE-FG02-04ER41268 and in part by a Humboldt Research Award sponsored by
the Alexander von Humboldt Foundation.
The work of J.P.S. is also funded by FCT through the projects
CERN/FP/109305/2009 and  U777-Plurianual,
and by the EU RTN project Marie Curie: PITN-GA-2009-237920.
H.E.H. is grateful for the hospitality of the Centro de F\'{\i}sica
Te\'{o}rica e Computacional, Faculdade de Ci\^{e}ncias, Universidade de Lisboa,
where this work was conceived and the Bethe Center for
Theoretical Physics at the Physikalisches Institut der Universit\"at
Bonn, where part of this work was completed.
}
%
%
%


\begin{thebibliography}{99}


\bibitem{ATLASHiggs}
G.~Aad {\it et al.}  [ATLAS Collaboration],
  Phys.\ Lett.\ B {\bf 716}, 1 (2012)
  [arXiv:1207.7214 [hep-ex]].

\bibitem{CMSHiggs}
S.~Chatrchyan {\it et al.}  [CMS Collaboration],
  Phys.\ Lett.\ B {\bf 716}, 30 (2012)
  [arXiv:1207.7235 [hep-ex]].

\bibitem{hhg}
  J.F.~Gunion, H.E.~Haber, G.L.~Kane and S.~Dawson,
  \textit{The Higgs Hunter's Guide}
  \mbox{(Westview Press, Boulder, CO, 2000)}.

\bibitem{ATLASgg}
ATLAS Collaboration, ATLAS-CONF-2012-168
(\texttt{https://atlas.web.cern.ch/Atlas/GROUPS/PHYSICS/CONFNOTES/ATLAS-}\\
\texttt{CONF-2012-168/}).


\bibitem{CMSgg}
CMS Collaboration, CMS-PAS-HIG-12-015
(\texttt{http://cdsweb.cern.ch/record/1460419?ln=en}).

\bibitem{manyrefs1}
M.~Carena, S.~Gori, N.R.~Shah and C.E.M.~Wagner,
  JHEP {\bf 1203}, 014 (2012)
  [arXiv:1112.3336 [hep-ph]];
A.~Arhrib, R.~Benbrik, M.~Chabab, G.~Moultaka and L.~Rahili,
  JHEP {\bf 1204}, 136 (2012)
  [arXiv:1112.5453 [hep-ph]];
J.-J.~Cao, Z.-X.~Heng, J.M.~Yang, Y-M.~Zhang and J.-Y.~Zhu,
  JHEP {\bf 1203}, 086 (2012)
  [arXiv:1202.5821 [hep-ph]];
A.~Arhrib, R.~Benbrik, M.~Chabab, G.~Moultaka and L.~Rahili,
  arXiv:1202.6621 [hep-ph];
M.~Carena, S.~Gori, N.R.~Shah, C.E.M.~Wagner and L.-T.~Wang,
  JHEP {\bf 1207}, 175 (2012)
  [arXiv:1205.5842 [hep-ph]];
A.G.~Akeroyd and S.~Moretti,
  Phys.\ Rev.\ D {\bf 86}, 035015 (2012)
  [arXiv:1206.0535 [hep-ph]];
C.-W.~Chiang and K.~Yagyu,
  arXiv:1207.1065 [hep-ph];
H.~An, T.~Liu and L.~-T.~Wang,
  Phys.\ Rev.\ D {\bf 86}, 075030 (2012)
  [arXiv:1207.2473 [hep-ph]];
A.~Joglekar, P.~Schwaller and C.E.M.~Wagner,
  JHEP {\bf 1212}, 064 (2012)
  [arXiv:1207.4235 [hep-ph]];
  B.~Batell, D.~McKeen and M.~Pospelov,
  JHEP {\bf 1210}, 104 (2012)
  [arXiv:1207.6252 [hep-ph]];
G.F.~Giudice, P.~Paradisi and A.~Strumia,
  JHEP {\bf 1210}, 186 (2012)
  [arXiv:1207.6393 [hep-ph]];
T.~Abe, N.~Chen and H.-J.~He,
JHEP {\bf 1301}, 082 (2013);
M.~Chala,
  JHEP {\bf 1301}, 122 (2013)
  [arXiv:1210.6208 [hep-ph]];
A.~Urbano,
  arXiv:1208.5782 [hep-ph].
  
\bibitem{manyrefs2}
U.~Ellwanger,
  JHEP {\bf 1203}, 044 (2012)
  [arXiv:1112.3548 [hep-ph]];
A.~Arhrib, R.~Benbrik and C.-H.~Chen,
  arXiv:1205.5536 [hep-ph];
V.~Barger, M.~Ishida and W.~-Y.~Keung,
  arXiv:1207.0779 [hep-ph].


\bibitem{logan}
B.~Coleppa, K.~Kumar and H.E.~Logan,
 Phys.\ Rev.\ D {\bf 86}, 075022 (2012)
  [arXiv:1208.2692 [hep-ph]].


\bibitem{sumrule}
J.F.~Gunion, H.E.~Haber and J.~Wudka,
  Phys.\ Rev.\ D {\bf 43}, 904 (1991).

\bibitem{sher}
P.M.~Ferreira, R.~Santos, M.~Sher and J.P.~Silva,
  Phys.\ Rev.\ D {\bf 85}, 077703 (2012)
  [arXiv:1112.3277 [hep-ph]].

\bibitem{2HDMrefs}
P.M.~Ferreira, R.~Santos, M.~Sher and J.P.~Silva,
  Phys.\ Rev.\ D {\bf 85}, 035020 (2012)
  [arXiv:1201.0019 [hep-ph]];
A.~Arhrib, R.~Benbrik and N.~Gaur,
  Phys.\ Rev.\ D {\bf 85}, 095021 (2012)
  [arXiv:1201.2644 [hep-ph]];
H.S.~Cheon and S.K.~Kang,
  arXiv:1207.1083 [hep-ph];
W.~Altmannshofer, S.~Gori and G.D.~Kribs,
   Phys.\ Rev.\ D {\bf 86}, 115009 (2012)
  [arXiv:1210.2465 [hep-ph]];
S.~Chang, S~K.~Kang, J.-P.~Lee, K.Y.~Lee, S.C.~Park and J.~Song,
  arXiv:1210.3439 [hep-ph];
Y.~Bai, V.~Barger, L.L.~Everett and G.~Shaughnessy,
  arXiv:1210.4922 [hep-ph].


 \bibitem{Gunion:2012gc}
  J.F.~Gunion, Y.~Jiang and S.~Kraml,
  Phys.\ Rev.\ D {\bf 86}, 071702 (2012)
  [arXiv:1207.1545 [hep-ph]].

\bibitem{degenerate2}
B.~Grzadkowski, ``Scalar-sector extensions in light of the LHC data,''
talk given at the Workshop on Multi-Higgs Models,
Complexo Interdisciplinar da UL, Lisbon, Portugal, 28--31 August 2012 (unpublished);
A.~Drozd, B.~Grzadkowski, J.F.~Gunion and Y.~Jiang,
  arXiv:1211.3580 [hep-ph].

\bibitem{Gunion:2012he}
J.F.~Gunion, Y.~Jiang and S.~Kraml,
  Phys.\ Rev.\ Lett.\  {\bf 110}, 051801 (2013)
  [arXiv:1208.1817 [hep-ph]].

\bibitem{review}
G.C.~Branco, P.M.~Ferreira, L.~Lavoura, M.N.~Rebelo, M.~Sher and J.P.~Silva,
  Phys.\ Rept.\  {\bf 516}, 1 (2012)
  [arXiv:1106.0034 [hep-ph]].

\bibitem{type1}
H.E.~Haber, G.L.~Kane and T.~Sterling,
Nucl.\ Phys.\ {\bf B161}, 493 (1979).


\bibitem{hallwise}
L.J.~Hall and M.B.~Wise,
Nucl.\ Phys.\ {\bf B187}, 397 (1981).


\bibitem{type2}
J.F.~Donoghue and L.F.~Li,
Phys.\ Rev.\ {\bf D19}, 945 (1979).

%
\bibitem{Spira:1995mt}
M.~Spira,
arXiv:hep-ph/9510347.

\bibitem{LHCHiggs}
S.~Dittmaier, C.~Mariotti, G.~Passarino, R.~Tanaka (editors) {\it et al.} [LHC Higgs Cross Section Working Group],
\textit{Handbook of LHC Higgs Cross Sections: 1. Inclusive Observables}, CERN Yellow Report, CERN-2011-002
[arXiv:1101.0593 [hep-ph]]; \textit{Handbook of LHC Higgs Cross Sections: 2. Differential Distributions}, CERN Yellow Report, CERN-2012-002 [arXiv:1201.3084 [hep-ph]].  The most recent
updates can be found on the LHC Higgs Cross Section Working Group webpage,
\texttt{https://twiki.cern.ch/twiki/bin/view/LHCPhysics/CrossSections}.

%
\bibitem{Harlander:2003ai}
  R.V.~Harlander and W.B.~Kilgore,
  Phys.\ Rev.\ D {\bf 68}, 013001 (2003)
  [hep-ph/0304035].
%



\bibitem{vac1}
  N.G.~Deshpande and E.~Ma,
  Phys.\ Rev.\  D {\bf 18} (1978) 2574.


\bibitem{unitarity}
S.~Kanemura, T.~Kubota and E.~Takasugi,
Phys.\ Lett.\  B {\bf 313} (1993)  155; 
A.G.~Akeroyd, A.~Arhrib and E.M.~Naimi,
  Phys.\ Lett.\  B {\bf 490} (2000)  119.

\bibitem{Peskin:1991sw}
  M.E.~Peskin and T.~Takeuchi,
  Phys.\ Rev.\ D {\bf 46}, 381 (1992).

\bibitem{STHiggs}
 H.E.~Haber,
  ``Introductory Low-Energy Supersymmetry,'' in
\textit{Recent directions in particle theory: from superstrings and black holes to the standard model}, Proceedings of
the Theoretical Advanced Study Institute (TASI 92), Boulder, CO, 1--26
June 1992, edited by J.~Harvey and J.~Polchinski (World Scientific
Publishing, Singapore, 1993) pp.~589--688;
  C.~D.~Froggatt, R.~G.~Moorhouse and I.~G.~Knowles,
  Phys.\ Rev.\  D {\bf 45}, 2471 (1992);
 W.~Grimus, L.~Lavoura, O.~M.~Ogreid and P.~Osland,
  Nucl.\ Phys.\ B {\bf 801}, 81 (2008)
  [arXiv:0802.4353 [hep-ph]];
  H.E.~Haber and D.~O'Neil,
  Phys.\ Rev.\ D {\bf 83}, 055017 (2011)
  [arXiv:1011.6188 [hep-ph]].

\bibitem{lepewwg}
S.~Schael et al. [The ALEPH, DELPHI, L3 and OPAL Collaborations, and the LEP
Electroweak Working Group], arXiv:1302.3415 [hep-ex], submitted to Physics
Reports.

\bibitem{gfitter1}
  M.~Baak, M.~Goebel, J.~Haller, A.~Hoecker, D.~Ludwig, K.~Moenig, M.~Schott and J.~Stelzer,
  Eur.\ Phys.\ J.\ C {\bf 72}, 2003 (2012)
  [arXiv:1107.0975 [hep-ph]].

\bibitem{gfitter2}
M.~Baak, M.~Goebel, J.~Haller, A.~Hoecker, D.~Kennedy, R.~Kogler, K.~Moenig, M.~Schott and J.~Stelzer,
  Eur.\ Phys.\ J.\ C {\bf 72}, 2205 (2012)
  [arXiv:1209.2716 [hep-ph]].

\bibitem{decoupling}
 J.F.~Gunion and H.E.~Haber,
  Phys.\ Rev.\ D {\bf 67}, 075019 (2003)
  [hep-ph/0207010].


\bibitem{Lees:2012xj}
  J.P.~Lees {\it et al.}  [BaBar Collaboration],
  Phys.\ Rev.\ Lett.\  {\bf 109}, 101802 (2012)
  [arXiv:1205.5442 [hep-ex]].

\bibitem{Freitas:2012sy}
  A.~Freitas and Y.-C.~Huang,
  JHEP {\bf 1208}, 050 (2012)
  [arXiv:1205.0299 [hep-ph]].

\bibitem{Denner:1991ie}
  A.~Denner, R.J.~Guth, W.~Hollik and J.H.~Kuhn,
  Z.\ Phys.\ C {\bf 51}, 695 (1991).

\bibitem{Boulware:1991vp}
  M.~Boulware and D.~Finnell,
  Phys.\ Rev.\ D {\bf 44}, 2054 (1991).

\bibitem{Grant:1994ak}
  A.K.~Grant,
  Phys.\ Rev.\ D {\bf 51}, 207 (1995)
  [hep-ph/9410267].

\bibitem{Haber:1999zh}
  H.E.~Haber and H.E.~Logan,
  Phys.\ Rev.\ D {\bf 62}, 015011 (2000)
  [hep-ph/9909335].

\bibitem{Misiak:2006zs}
  M.~Misiak {\it et al.},
  Phys.\ Rev.\ Lett.\  {\bf 98}, 022002 (2007)
  [hep-ph/0609232].

\bibitem{Asner:2010qj}
Y.~Amhis et al. [Heavy Flavor Averaging Group (HFAG)]
arXiv: 1207.1158 [hep-ex].

\bibitem{Mahm}
T.~Hermann, M.~Misiak and M.~Steinhauser,
  JHEP {\bf 1211} (2012) 036
  [arXiv:1208.2788 [hep-ph]]. See also
%
F. Mahmoudi, talk given at Prospects For Charged Higgs Discovery At Colliders
(CHARGED 2012), 8-11 October, Uppsala, Sweden.

\bibitem{BB}
  F.~Mahmoudi and O.~Stal,
  Phys.\ Rev.\ D {\bf 81}, 035016 (2010)
  [arXiv:0907.1791 [hep-ph]];
  S.~Su and B.~Thomas,
  Phys.\ Rev.\ D {\bf 79}, 095014 (2009)
  [arXiv:0903.0667 [hep-ph]];
  M.~Aoki, S.~Kanemura, K.~Tsumura and K.~Yagyu,
  Phys.\ Rev.\ D {\bf 80}, 015017 (2009)
  [arXiv:0902.4665 [hep-ph]];
 P.~Posch,
University of Vienna Ph.D. dissertation (2009).

\bibitem{bsgmssm}
 R.~Barbieri and G.F.~Giudice,
  Phys.\ Lett.\ B {\bf 309}, 86 (1993)
  [hep-ph/9303270].


\bibitem{Barroso:2012wz}
  A.~Barroso, P.M.~Ferreira, R.~Santos and J.P.~Silva,
  Phys.\ Rev.\ D {\bf 86}, 015022 (2012)
  [arXiv:1205.4247 [hep-ph]].

\bibitem{mssmhiggs}
S.~Schael et al. [ALEPH and DELPHI and L3 and OPAL Collaborations and the LEP
Working Group for Higgs Boson Searches], Eur.\ Phys.\ J.\ C {\bf 47}, 547
(2006).


\end{thebibliography}
\end{document}